\documentclass[a4paper,11pt]{article}
\usepackage{jheppub}
\usepackage[T1]{fontenc}
\usepackage{tikz-feynman}
\usepackage{framed}
\usepackage[capitalise]{cleveref}
\usepackage{graphicx}
\usepackage{subcaption}

\newcommand{\gsim}{\;\rlap{\lower 3.5 pt \hbox{$\mathchar \sim$}} \raise 1pt
\hbox {$>$}\;}
\newcommand{\lsim}{\;\rlap{\lower 3.5 pt \hbox{$\mathchar \sim$}} \raise 1pt
\hbox {$<$}\;}

\newcommand{\ORD}[1]{\mathcal{O}\left(#1\right)}
\newcommand{\VEC}[1]{\boldsymbol{#1}}
\newcommand{\Li}[2]{\text{Li}_{#1}(#2)}
\newcommand{\Lifoh}{\Li{4}{1/2}}
\def\me{m_{e}}
\def\mE{m_{\mu}}
\def\ma{m_{\gamma}}
\def\mesq{m_{e}^2}
\def\mEsq{m_{\mu}^2}
\def\masq{m_{\gamma}^2}
\def\be{\beta_{e}}
\def\bE{\beta_{\mu}}
\def\ne{n_{e}}
\def\nE{n_{\mu}}
\def\nl{n_l}
\def\nF{n_F}
\def\nFp{n_{F'}}
\def\nFb{n_{\overline{F}}}

\def\LEe{L_{\mu e}}
\def\abare{\alpha_{\mathrm{b}}}
\def\varBUE{v}
\def\varBUeE{w}
\def\betafrac{z_\beta}
\def\ALL{A^{\text{LL}}}
\def\ANLL{A^{\text{NLL}}}

\title{\boldmath Electron-to-Muon Forward Annihilation at High Energy}

\preprint{\parbox[t]{2in}{CERN-TH-2026-264\\ TUM-HEP-1618/26 }}

\author[a]{Maximilian Delto,}
\author[b]{Alexander A. Penin,}
\author[c]{Lorenzo Tancredi\,}
\affiliation[a]{Theoretical Physics Department, CERN,  CH-1211 Geneva 23, Switzerland}
\affiliation[b]{Department of Physics, University of Alberta,
Edmonton AB T6G 2E1, Canada}
\affiliation[c]{TUM School of Natural Sciences, Physik Department, James-Franck-Straße 1,
Technische Universität München, D--85748 Garching, Germany}

\emailAdd{maximilian.delto@cern.ch}
\emailAdd{penin@ualberta.ca}
\emailAdd{lorenzo.tancredi@tum.de}

\abstract{We study the  high-energy  behavior of
electron-to-muon annihilation in forward kinematics in QED
with two massive lepton species.
Using effective field theory renormalization group
techniques we derive the all-order asymptotic
behavior of the annihilation amplitude
through the next-to-leading logarithmic approximation.
Furthermore, we consider one- and two-loop
fixed-order corrections to the forward-annihilation amplitude
in the high-energy limit, which we compute
up to power corrections. The goal of this paper is
to elaborate on various technical details that
we omitted in the original presentation of our
results in Ref.~\cite{Delto:2025epy}.
}

\begin{document}
\maketitle
\flushbottom

\section{Introduction}
\label{sec::int}
Electron-positron annihilation into a muon-antimuon pair at
high energy is a  benchmark process that has been used for
precision tests of quantum electrodynamics (QED) and the
electroweak theory.  The required high precision of
theoretical predictions makes the calculation of radiative
corrections to the process cross section indispensable. 
While next-to-leading order studies were completed in the early days of the Standard Model
\cite{Putzolu:1961jya,Berends:1973tz,Berends:1973fd,Passarino:1978jh},
the two-loop QED corrections have been obtained only recently
\cite{Bonciani:2021okt} following a significant advance 
in multi-loop methods. In this work, we consider  the high-energy
limit where the ratios of the fermion masses $m_{e,\mu}$ to the
center-of-mass energy $\sqrt{s}$ are small. The forward  limit
corresponds to small scattering angles $\theta\lsim
m_\mu^2/s$ when the total momentum transfer between the same
electric charges  scales as $(m_\mu^2-m_e^2)/\sqrt{s}$. Although
the contribution of the forward region to the  annihilation
cross section is power-suppressed by the final state phase
space, this kinematical limit is of significant theoretical
interest. The available two-loop results
\cite{Bonciani:2021okt,Broggio:2022htr} obtained through infrared matching
\cite{Penin:2005kf,Penin:2005eh,Bonciani:2007eh,Bonciani:2008ep}
cannot be extended to the forward region because of its
peculiar infrared structure which, in particular, results in a
highly nontrivial asymptotic behavior of the annihilation
amplitude in  this
regime~\cite{Gorshkov:1966ht,Gorshkov:1973if}. The  asymptotic
behavior  is governed by the radiative corrections enhanced by
powers of large logarithms of the scale ratio,  which
invalidate fixed-order perturbation theory and should be
resummed to all orders in the fine structure constant $\alpha$.
In contrast to the wide-angle Sudakov regime~\cite{Sudakov:1954sw},
where the logarithmic corrections are associated with soft
gauge-boson exchange, in the forward regime the dominant
logarithmic contributions arise from configurations involving
soft fermions. While the analysis of the Sudakov limit
and the small-angle Regge 
limit~\cite{Regge:1959mz,Regge:1960zc,Frolov:1970ij,Cheng:1970xm}
has been the focus of research for a long time and has been
generalized to non-Abelian gauge groups and subleading
logarithms, little progress has been achieved for the
forward kinematics, which is much less understood.
In particular, this is related to the highly nontrivial origin and
structure of the logarithmically enhanced  terms, which make
the problem quite challenging for the available theoretical
techniques and very interesting in the general context of
effective field theory. 
Although in recent years 
many nontrivial results have been obtained in the
renormalization group analysis of mass-suppressed
amplitudes by different methods (see
\cite{Laenen:2010uz,Penin:2014msa,Melnikov:2016emg,
Bonocore:2016awd,Penin:2016wiw,Liu:2017axv,Liu:2017vkm,
Boughezal:2016zws,Moult:2018jjd,Liu:2018czl,Beneke:2018gvs,
Ebert:2018gsn,Bahjat-Abbas:2019fqa,Beneke:2019mua,
Moult:2019uhz,Liu:2019oav,Wang:2019mym,Anastasiou:2020vkr,Liu:2020wbn,
Penin:2019xql,Beneke:2022obx,Liu:2021chn,Liu:2022ajh,
AH:2020iki,Liu:2024tkc} for a few examples) for more than six
decades the theory of forward annihilation or, 
equivalently, of electron-muon backward scattering, has been limited 
to the original diagrammatic resummation of the leading double
logarithmic terms \cite{Gorshkov:1966ht}. Only recently has the
problem  been discussed from the effective field theory
point of view \cite{Bell:2022ott,Bell:2024bxg}.

In Ref.~\cite{Delto:2025epy} we introduced a new approach
to the renormalization group analysis of power-suppressed
amplitudes. It provides a powerful and adaptive tool,
applicable to  a  wide range of processes and kinematical
regimes with a diverse spectrum of physical modes, from soft
fermions to Glauber gauge bosons, which  pose a serious
challenge to existing techniques. The method has been used, in
particular, for the analysis of electron-to-muon forward
annihilation in the high-energy limit. The aim of this paper is
to give a detailed account of this analysis. It is structured
as follows. In \cref{sec::notation-kinematics}, we give a
precise definition of the forward-annihilation process and
parametrize its kinematics. We then analyze one-loop QED
corrections to the forward-annihilation process for equal and
unequal lepton masses in \cref{sec::one-loop}. In
particular, we provide the analysis of the regions and modes
that generate logarithmically enhanced corrections in the
forward regime. Building upon this understanding, in
\cref{sec::evolution}, we discuss factorization and
effective-theory renormalization group flow, and derive the
asymptotic behavior of the amplitude through the
next-to-leading logarithmic approximation. In
\cref{sec::2LFO}, we describe the calculation of fixed-order
two-loop corrections in the high-energy forward limit in the
cases of equal and different lepton masses. We conclude in \cref{sec::conclusion}.

\section{Notation and kinematics}
\label{sec::notation-kinematics}
We study electron-to-muon annihilation
\begin{equation}
\label{eqn:def-process}
e^+(p_1) + e^-(p_2) \to  \mu^+(p_3) + \mu^-(p_4) \,,
\end{equation}
for massive leptons $p_{1,2}^2=\mesq$ and $p_{3,4}^2=\mEsq$.
We work in the strict forward limit $\theta_{13}=\theta_{24}=0$,
which gives the leading approximation in the forward
annihilation regime corresponding to $\theta_{13,24}\ll m_\mu^2/s$.
Then, in the center-of-mass frame, we can define the momenta
\begin{align}
\label{eqn:p1234-FS-COMF}
p_{1,2} =  \frac{\sqrt{s}}{2}(1, \pm \be \VEC{n} ) ~,~ \quad p_{3,4} 
=  \frac{\sqrt{s}}{2}(1, \pm \bE \VEC{n} ) \,,
\end{align}
in terms of the lepton velocities $\beta_x=\sqrt{1-4m_x^2/s}$, the spatial
directions of flight $\VEC{n}^2=1$, and the center-of-mass
energy squared $s=(p_1+p_2)^2$. 
To properly set up the power-counting in the
high-energy limit, we define a small expansion parameter
$\lambda = \mE / \sqrt{s} \ll 1$ and the mass ratio $x = \me  /
\mE \in [0,1]$. While we will keep lepton flavors distinct
throughout the paper, we will consider two cases for lepton
masses: the physically motivated case,
where the electron is much lighter than the muon
($\mesq\ll \mEsq \ll s,\, x \ll 1$), and the more formal
case where both fermion masses are of equal magnitude
and small compared to the center-of-mass energy
($\mesq \sim \mEsq \ll s,\, x \sim 1$). We will refer to these
cases as the \textit{ordered
masses} (OM) and the \textit{equal masses} (EM) cases, respectively.
In the high-energy limit the
logarithm
\begin{align}
    \label{eqn:def-L}
    L
    =\ln \left(  \frac{s}{\mEsq} \right)
    =\ln \left( \frac1{\lambda^2} \right) \,,
\end{align}
becomes large. Its allowed parametric scaling depends on the
approximation. The  leading logarithmic (LL) approximation,
which sums up the {\it double logarithmic} terms of the form
$\alpha^n L^{2n}$, is valid up to energies
for which $L\sim 1/\alpha^{1/2}$. The
next-to-leading logarithmic (NLL) approximation, which accounts
for the terms $\alpha^n L^{2n-1}$, is valid for $L\lsim
1/\alpha$, and so on. Note that in the case of ordered masses,
the radiative corrections include the energy-independent
logarithm of the mass ratio
\begin{align}
    \label{eqn:def-Lmue}
    \LEe 
    =\ln \left( \frac{\mEsq}{\mesq} \right) 
    =\ln \left( \frac1{x^2} \right) \,.
\end{align}
Numerically $\LEe\approx 10\ll 1/\alpha$ and we do \textit{not}
perform the resummation of the corresponding terms, which
is a separate problem. In particular, we will
provide results for the OM case $x\ll1$ 
disregarding power corrections,
and for the EM case in the regular limit $x=1$.

Making use of the forward kinematics, we define light-cone
directions $n_\pm=(1,\pm\VEC{n})$ with $n_+\cdot n_-=2$ and
$n_-^2=n_+^2=0$, and the decomposition
\begin{equation}
    \ell^\nu =   \frac{\ell^-}{{2}}n_+^\nu +  \frac{\ell^+ }{{2}}n_-^\nu
    + \ell_\perp \, e_\perp^\nu\,, \;\; n_\pm \cdot e_\perp=0 \,, 
    \;\; \ell^\pm = {n_\pm \cdot \ell } \,.
    \label{eqn:def-LCD}
\end{equation}
In this frame $e_\perp$ is a spacelike unit vector
$e_\perp^2=-1$.
External momenta in the light-cone decomposition in \cref{eqn:def-LCD} read
\begin{align}
    p_1^\pm = p_2^\mp =  \sqrt{s} {1 \mp \be \over 2}  ~~~\,,~~~ 
    p_3^\pm = p_4^\mp = \sqrt{s} {1 \mp \bE \over 2}   \,,~~~ p_{i,\perp}=0 \,.
\end{align}
By construction, their high-energy behavior yields 
power-suppressed components
\begin{align}
    p_1^+ = p_2^- \approx \sqrt{s}\, x^2 \lambda^2 \,, 
    ~~~~ p_3^+ = p_4^- \approx \sqrt{s} \, \lambda^2 \,,
\end{align}
and hard ones
\begin{align}
    p_1^- = p_2^+ = p_3^- = p_4^+ \approx \sqrt{s}\,.
\end{align}
The momentum transfer reads
\begin{align}
    q^\nu \equiv p_3^\nu-p_1^\nu = p_2^\nu - p_4^\nu = 
    \frac{\sqrt{s}}{4} (\be - \bE)(n_- - n_+)^\nu\,,
\end{align}
such that
\begin{align}
q^\pm = \pm\sqrt{s} {\be -\bE\over 2} \approx \pm \sqrt{s} \, \lambda^2 (1-x^2) \,,
\end{align}
and
\begin{align}
    t = q^2 = q^+ q^-
    = - \frac{s}{4} (\be-\bE)^2
    \approx - s \lambda^4 (1-x^2)^2 \,.
\end{align}

In the following section we will study how the amplitude and
the cross section behave in the kinematic limits defined above.
We write the perturbative series for the amplitude as
\begin{equation}
{\cal M}=\sum_{n=0}^\infty\left(\frac{\abare}{2\pi}\right)^n{\cal M}^{(n)} \,,
\end{equation}
where ${\cal M}^{(0)}={\cal M}_{\rm Born}$
and $\abare$ is the bare coupling constant. 
Later on, we will also need the renormalized value of the
coupling constant. We use $\bar{\alpha}(\mu)$ to indicate
the renormalized coupling constant in the $\overline{\text{MS}}$
scheme at scale $\mu$, and $\alpha$ for the corresponding
one in the on-shell scheme. As the logarithms in the running
do not depend on the scheme up to the two-loop order due to
the universality of the first two coefficients of the $\beta$
function, with some abuse of notation we will use $\alpha(\mu)$
to indicate the running of the coupling constant in a generic
scheme, having in mind the evolution of the physical effective
charge $\alpha(\mu) = \alpha_{\text{eff}}(\mu)$.

\section{Logarithmic behavior at one loop}
\label{sec::one-loop}

\begin{figure}[t]
  \centering
  \begin{subfigure}[b]{2.6cm}
    \centering
    \includegraphics[width=2.6cm]{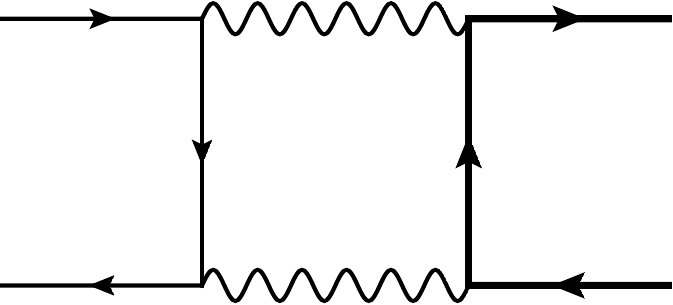}
    \caption{}\label{fig::1a}
  \end{subfigure}\hspace{2mm}
  \begin{subfigure}[b]{2.6cm}
    \centering
    \includegraphics[width=2.6cm]{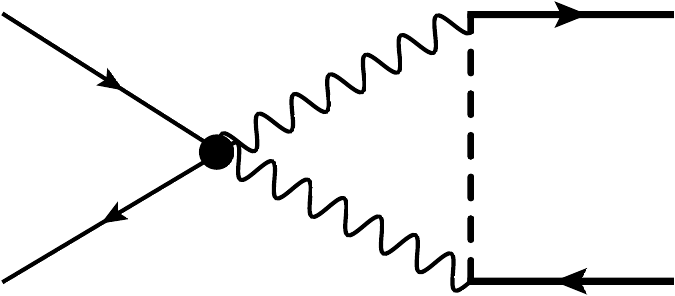}
    \caption{}\label{fig::1b}
  \end{subfigure}\hspace{2mm}
  \begin{subfigure}[b]{2.6cm}
    \centering
    \includegraphics[width=2.6cm]{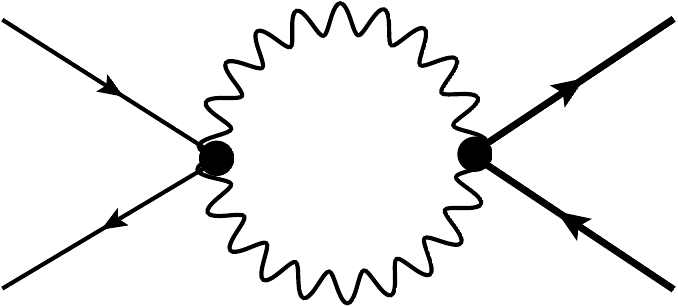}
    \caption{}\label{fig::1c}
  \end{subfigure}\hspace{2mm}
  \begin{subfigure}[b]{2.6cm}
    \centering
    \includegraphics[width=2.6cm]{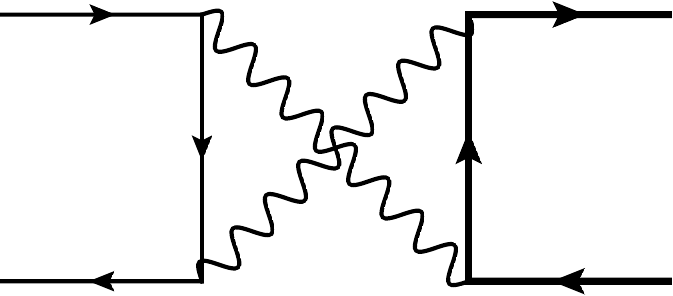}
    \caption{}\label{fig::1d}
  \end{subfigure}\hspace{2mm}
  \begin{subfigure}[b]{2.0cm}
    \centering
    \includegraphics[width=2.0cm]{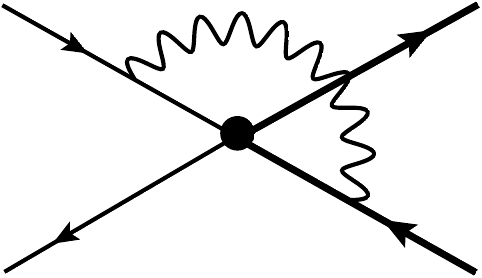}
    \caption{}\label{fig::1e}
  \end{subfigure}
  \caption{The full and reduced one-loop box
  diagrams generating the double and single logarithmic
  corrections. The bold (thin) fermion line corresponds to the
  muon (electron) field. The black  circles represent the
  effective vertices from collapsed fermion or photon
  propagators. The bold dashed line is the effective scalar
  muon propagator.}
  \label{fig::1}
\end{figure}

We start with the logarithmic behavior of the amplitude at
one loop. The vertex, vacuum polarization, and  self-energy
corrections are well known and proportional to the Born
amplitude. The nontrivial contribution comes from the box
diagram and its crossed counterpart, see \cref{fig::1a}
and \cref{fig::1d}, respectively.
In Feynman gauge the box diagram reads
\begin{align}
    \label{eqn:diag-box}
    & i\mathcal{M}^{(1)}_{a} =(-ie)^4 \int \frac{{\rm d}^4 \ell}{(2\pi)^4} \,\\
    &\frac{ \left[ \overline{V}_e(p_1) \, \gamma^\alpha ( \widehat{\ell} + \me ) \gamma^\beta \, U_e(p_2) \right]
    \,
    \left[ \overline{U}_\mu(p_4) \, \gamma_\beta  (\widehat{\ell} - \widehat{q} + \mE )  \gamma_\alpha  \, V_\mu(p_3) \right]
    }{
    \left[\ell^2 - \mesq \right]
    \left[(\ell+p_1)^2 \right]
    \left[(\ell-q)^2 - \mEsq \right]
    \left[(\ell-p_2)^2 \right]
    } \nonumber \,,
\end{align}
where $\widehat{v} \equiv v^\mu \gamma_\mu$.
The crossed box is given by
\begin{align}
    \label{eqn:diag-cross-box}
    & i\mathcal{M}^{(1)}_{d} = (-ie)^4\int \frac{{\rm d}^4 \ell}{(2\pi)^4} \\
    & \frac{ \left[ \overline{V}_e(p_1) \, \gamma^\alpha ( -\widehat{\ell}  -\widehat{p}_1  + \me ) \gamma^\beta \, U_e(p_2) \right] \, \left[ \overline{U}_\mu(p_4) \, \gamma_\alpha  (\widehat{\ell} + \widehat{p}_4 + \mE )  \gamma_\beta  \, V_\mu(p_3) \right] } {\left[(\ell+p_1)^2 - \mesq \right] \left[\ell^2 \right] \left[(\ell+p_4)^2 - \mEsq \right] \left[(\ell+p_1+p_2)^2 \right] } \nonumber \,.
\end{align}

\paragraph{Ordered masses.}
We begin by discussing the case $\me \ll \mE$. In this regime
we neglect the electron mass unless it serves as a collinear
regulator of the loop integral.  Due to the on-shell condition
for the external fermion lines, only the loop momentum's
transverse component contributes to the amplitude without
chirality flip, which is also the one we are interested
in.\footnote{The chirality changing amplitude starts to
contribute to the cross section only in the
next-to-next-to-leading logarithmic approximation. For this
reason we only consider the projection on the tensor structure
of the Born.} Thus the box diagram  reduces to the following
integral

\begin{equation}
\frac{\mathcal{M}^{(1)}_{a}}{{\cal M}_{\rm Born}}
\sim -i{Q^2\over \pi^2} \int{\ell_\perp^2  
{\rm d}^4\ell\over \ell^2(\ell^2-m_\mu^2)
(\ell+p_1)^2(\ell-p_2)^2}\,,
\label{eq::planarint}
\end{equation}
where $Q^2=-s - i 0$ and  we neglect the momentum transfer. Writing
\begin{equation}
\ell_\perp^2=-\ell^2+\ell^+\ell^-\, \label{eq::lperpsq}
\end{equation}
we can split the integral above in two contributions of
different logarithmic scaling. The $\ell^2$ term cancels the
electron propagator  and the corresponding contribution is
represented by the effective diagram in
\cref{fig::1b}
with the muon propagator  substituted by the  scalar one
$1/(\ell^2-m_\mu^2)$, which is sufficiently singular at small
momentum to develop a double logarithmic contribution
\begin{equation}
\frac{\mathcal{M}^{(1)}_{b}}{{\cal M}_{\text{Born}}}
= i{Q^2\over \pi^2}\int { {\rm d}^4\ell\over
(\ell^2-m_\mu^2)(\ell+p_1)^2(\ell-p_2)^2}=
{1\over 2}\ln^2\left({Q^2\over m_\mu^2}\right)+\ldots\,.
\label{eq::dlint}
\end{equation}
In the equation above and in the following, non-logarithmic
terms are disregarded as denoted by the ellipsis. This diagram
has a structure similar to the one-loop amplitude of fermion
scattering by a quadratic gauge field operator, where the
scalar part of the soft fermion propagator corresponds to the
chirality  flipping mass term
\cite{Liu:2017vkm,Anastasiou:2020vkr}. Following this analysis
we associate  this non-Sudakov double logarithmic term with the
contribution of the on-shell soft muon and off-shell eikonal
transversely polarized photons, which corresponds to the
following approximation for the propagators
\begin{equation}
{1\over \ell^2-m_\mu^2}  \approx - 2\pi i
\delta(\ell^2-m_\mu^2)\,,\quad
{1\over (\ell+p_1)^2}\approx \frac{1}{2\ell \cdot p_1}\,,\quad
{1\over (\ell-p_2)^2} \approx -\frac{1}{2\ell \cdot p_2}\,.
\label{eq::prop}
\end{equation}
With the logarithmic integration region
\begin{align}
{m_\mu^2\over \sqrt{s} }< \ell^{\pm} < \sqrt{s} \,,
\label{eqn:def-scaling-region}
\end{align}
and an additional constraint
\begin{align}
 \ell^+\ell^- > m_\mu^2  \,,
 \label{eqn:def-scaling-OS}
\end{align}
from the muon on-shell condition, this immediately reproduces
the double-logarithmic term in \cref{eq::dlint}.

The $\ell^+\ell^-$ term in the expression for $\ell_\perp^2$
given in \cref{eq::lperpsq} can be rewritten as follows
\begin{equation}
\ell^+\ell^-\approx -{4\over Q^2}
(\ell \cdot p_1)(\ell \cdot p_2)\approx
{1\over Q^2}\left((\ell+p_1)^2-\ell^2\right)
\left((\ell-p_2)^2-\ell^2\right)\,,
\label{eq::lplmdec}
\end{equation}
where only the $\ell^4$ term  gives an integral depending on
both $p_1$ and $p_2$ which can produce the logarithmic
dependence on $Q$. After  substituting  $\ell^+\ell^-\to
\ell^4/Q^2$  both  the electron and  muon propagators  cancel
up to mass suppressed terms and the corresponding contribution
is represented by the effective bubble diagram in
\cref{fig::1c}. In the integration region
\cref{eqn:def-scaling-region,eqn:def-scaling-OS} we find the
single-logarithmic enhancement
\begin{equation}
\frac{\mathcal{M}^{(1)}_{c}}{{\cal M}_{\rm
Born}} = -{i\over \pi^2}\int
{{\rm d}^4\ell\over (\ell+p_1)^2(\ell-p_2)^2}=
-\ln\left({Q^2\over m_\mu^2}\right)+\ldots\,.
\label{eq::slint}
\end{equation}
We note that the apparent ultraviolet divergent contributions
to the above integral cancel against the remaining  terms in
\cref{eq::lplmdec}.

The logarithmic contribution of the crossed box diagram,
\cref{fig::1d}, is quite different. It is generated from
configurations where the hard momentum $p_1+p_2$ flows through
one of the photon propagators. Such a propagator is therefore
far off-shell in the high-energy limit and collapses, producing
an effective four-fermion vertex, see \cref{fig::1e}, and its
symmetric counterpart. The corresponding logarithmic
enhancement is of the usual Sudakov type, and the corresponding 
contribution\footnote{We include a factor of two to account for 
the  symmetric contribution $p_{1,4} \to p_{2,3}$.} 
reads~\cite{Kuhn:1999nn}
\begin{align}
\label{eq::npbox}
\frac{2\mathcal{M}^{(1)}_{e}}{{\cal M}_{\rm Born}}={} &
-2i{Q^2\over \pi^2} \int 
\frac{{\rm d}^4\ell}{\left[(\ell+p_1)^2 -\mesq \right] 
\left[ \ell^2-\masq\right] \left[ (\ell+p_4)^2 - \mEsq \right] } 
\nonumber \\
={} & \ln^2\left({s\over \masq}\right)
+\ldots\,,
\end{align}
where $p_4\approx p_2$ in the forward limit, and $\ma$ denotes
a photon mass which we introduced to regulate the infrared
divergence. The double-logarithmic contribution in
\cref{eq::npbox} cancels against the double-logarithmic part of
the electron and muon Sudakov form factors, each of which gives
\begin{equation}
\frac{\mathcal{M}^{(1)}}{{\cal M}_{\rm
Born}} \bigg|_{\text{form factor}}= 
-{1\over 2}\ln^2\left({Q^2\over \masq}\right)+
\gamma_f^{(1)}\ln\left({Q^2\over m_{\mu,e}^2}\right)+\ldots\,,
\label{eq::trigs}
\end{equation}
where $\gamma^{(1)}_f=3/2$ is the collinear anomalous dimension
of the fermion field. The cancellation leaves behind
single-logarithmic terms, regulated by the corresponding
fermion mass and proportional to $\gamma_f^{(1)}$.

Putting everything together and renormalizing the charge at the
scale $\mu = \mE$ we get
\begin{align}
{\cal M}^{(1)} =&{}
\left[{1\over 2}L^2
+\left(2\gamma^{(1)}_f-{\beta_0\over 2}-1+i\pi\right)L +c^{(1)}_{\text{IR}}
\right]{\cal M}_{\rm Born}+ \ldots
\nonumber \\
=&{}
\left[{1\over 2}L^2
+\left(2+{2\over 3}n_l+i\pi\right)L+c^{(1)}_{\text{IR}}\right]
{\cal M}_{\rm Born} + \ldots
\,,
\label{eq::1l}
\end{align}
where  $\beta_0=-{4\over 3}n_l$ is the one-loop coefficient of
the QED beta function with the $n_l=2$ active flavors at the
scale $Q$, and we omitted terms which do not contribute to the
cross section in the next-to-leading logarithmic approximation.
However, in \cref{eq::1l} we keep the energy-independent
logarithms of the mass ratio through the coefficient
\begin{equation}
c^{(1)}_{\text{IR}} = \left[\left(L_{\mu e}-2\right)
\ln\left({\masq\over m_\mu m_e}\right)
+\gamma^{(1)}_fL_{\mu e}\right]\,,
\label{eq::1lIRlogs}
\end{equation}
which is  infrared divergent and depends on the regularization
scheme. Note that for the symmetric choice of the soft cutoff
$\masq=m_\mu m_e$ the first term in \cref{eq::1lIRlogs}
disappears and the dependence of the correction on the electron
mass is entirely  due to the collinear renormalization of the
on-shell electron lines. The imaginary part of \cref{eq::1l}
cancels out in  the differential cross section through the
next-to-leading logarithmic approximation and can be omitted.

\paragraph{Equal masses.}
An interesting feature of the next-to-leading logarithmic
contribution is its uncharacteristic sensitivity to the details
of the  fermion mass spectrum, that distinguishes the forward
regime from massive fixed-angle scattering
\cite{Kuhn:1999nn,Kuhn:2001hz,Feucht:2004rp,Jantzen:2005az,
Kuhn:2007ca,Kuhn:2011mh,Penin:2011aa}. For a finite electron
mass $\mesq \sim \mEsq$  the reduction of the box
diagram  requires the following modification  of  \cref{eq::lperpsq}
\begin{equation}
\ell_\perp^2=(-\ell^2+m_e^2)+\ell^+\ell^--m_e^2\, ,
\label{eq::lperpsqmod}
\end{equation}
where now the terms in the brackets cancel the massive electron
propagator and result in the effective diagram \cref{fig::1b}
discussed above. The last term in \cref{eq::lperpsqmod} gives
an additional contribution, proportional to the electron mass
squared. It can be transformed into a difference of two
double-logarithmic integrals with the different infrared scales
\begin{align}
\frac{\mathcal{M}^{(1)}_a}{{\cal M}_{\rm
Born}}\bigg|_{m_e^2} \sim {} &
i{Q^2\over \pi^2}
\int{m_e^2 \, {\rm d}^4 \ell \over (\ell^2-m_e^2)(\ell^2-m_\mu^2)
(\ell+p_1)^2(\ell-p_2)^2}
\nonumber\\
={}&i{Q^2\over \pi^2}
\int {m_e^2\over m_\mu^2-m_e^2}\left({1\over (\ell^2-m_\mu^2)}-
{1\over (\ell^2-m_e^2)}\right){{\rm d}^4 \ell \over
(\ell+p_1)^2(\ell-p_2)^2}
\nonumber\\
={}&-{m_e^2\over m_\mu^2-m_e^2}
\ln\left({m_\mu^2\over m_e^2}\right)
\ln\left({Q^2\over m_\mu m_e}\right)+\ldots\,.
\label{eq::mecont}
\end{align}
Including this correction changes the coefficient of the single
logarithmic term in \cref{eq::1l} as
\begin{align}
\left(2\gamma^{(1)}_f-\frac{\beta_0}{2}-1+i\pi\right)L \rightarrow
\left(2\gamma^{(1)}_f-\frac{\beta_0}{2}+\gamma^{(1)}_e-1+i\pi\right)L\,,
\label{eq:EM-1L-NLL-coeff}
\end{align}
where we introduced
\begin{equation}
\gamma^{(1)}_{e}=-\frac{m_{e}^2}{m_\mu^2-m_e^2}
\ln\left({m_\mu^2\over m_e^2}\right)\,.
\label{eq::memu}
\end{equation}
In the strict equal-mass limit $m_e=m_\mu$  it reduces to
$\gamma^{(1)}_{e}  =-1$, and the one-loop contribution reads
\begin{align}
{\cal M}^{(1)} =
\left[{1\over 2}L^2+\left(1+{2\over 3}n_l+i\pi\right)L\right]
{\cal M}_{\rm Born} + \ldots\,,
\label{eq::1leqm}
\end{align}
where we omitted terms beyond NLL approximation.

\section{High-energy evolution in the  forward limit}
\label{sec::evolution}

In forward kinematics  the velocities of the electric
charges of the initial and final states are approximately
unchanged and soft radiation is strongly suppressed. Hence the
standard Sudakov suppression is absent. On the contrary, the
amplitude develops characteristic non-Sudakov radiative
corrections enhanced by the second power of the large logarithm
$L$ for each power of the fine structure constant $\alpha$,
resulting in the asymptotic {\em enhancement} of the  amplitude
at high energy \cite{Gorshkov:1966ht,Gorshkov:1973if}.
In \cite{Delto:2025epy} we introduced a
new class of evolution equations which govern the high
energy behavior of the power-suppressed contributions in
various kinematical limits. These equations describe the change
of the amplitudes under the variation of the effective-theory
cutoff and, therefore, can be understood as a Wilsonian
renormalization group flow. The main difference with respect to
the existing analysis is the use of more than one cutoff
dictated by the factorization structure and the geometry of the
effective-theory modes. This yields a system of, in general,
second-order partial differential equations. By solving this
system and removing the auxiliary cutoffs the  logarithmic
corrections are resummed to all orders in the coupling
constant. Below we describe how this strategy applies to the
forward annihilation amplitude.

\begin{figure}[t]
  \centering
  \begin{subfigure}[b]{2.6cm}
    \centering
    \includegraphics[width=2.6cm]{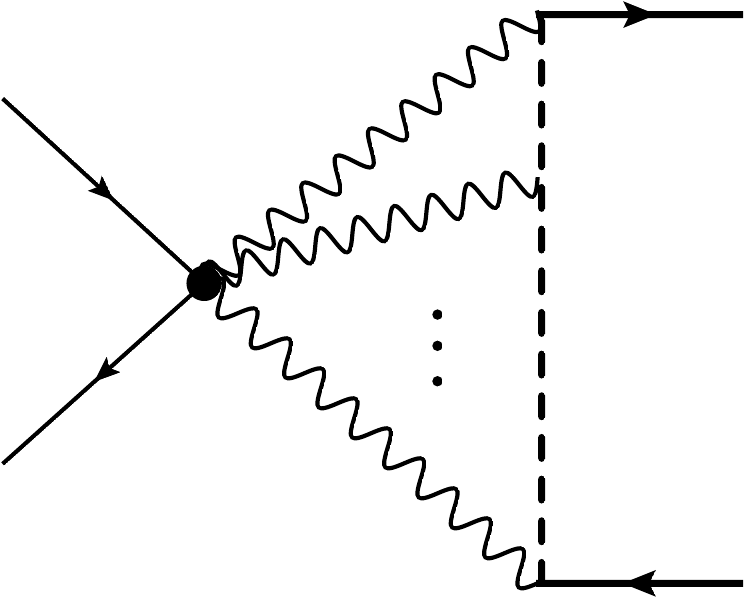}
    \caption{}\label{fig::2a}
  \end{subfigure}\hspace{10mm}
  \begin{subfigure}[b]{2.6cm}
    \centering
    \includegraphics[width=2.6cm]{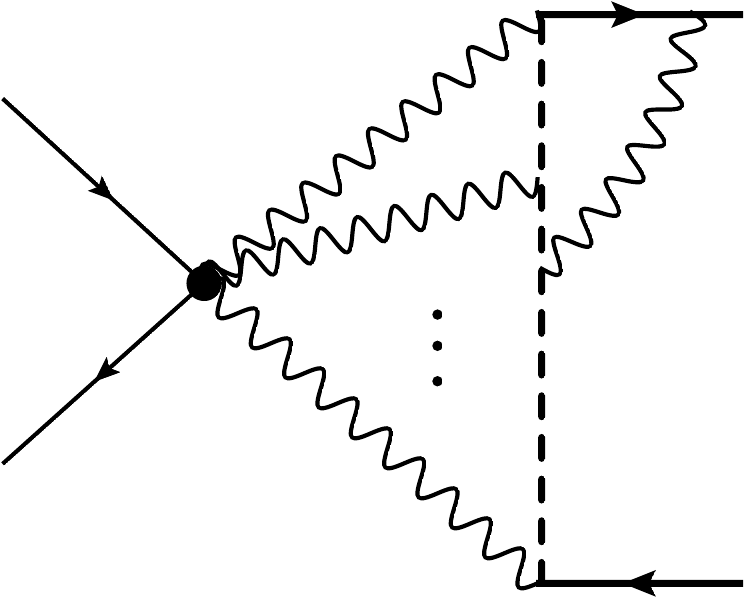}
    \caption{}\label{fig::2b}
  \end{subfigure}\hspace{10mm}
  \begin{subfigure}[b]{2.6cm}
    \centering
    \includegraphics[width=2.6cm]{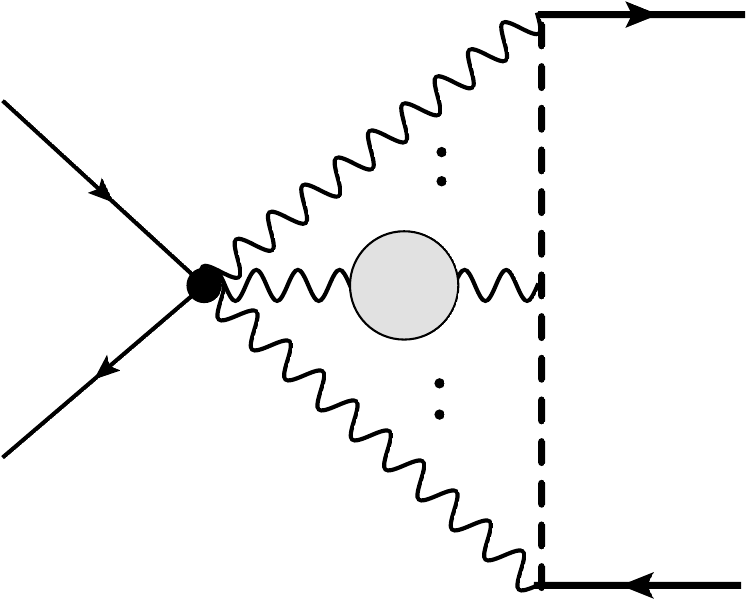}
    \caption{}\label{fig::2c}
  \end{subfigure}
  \caption{(a) The effective all-order Feynman
  diagrams representing  the leading logarithmic contribution.
  (b) An example of a topology not contributing in the
  next-to-leading logarithmic approximation. (c) The effective
  all-order  Feynman diagrams corresponding to the next-to-leading
  logarithmic contribution.  The gray circle represents the
  one-loop soft kernel \cref{eq::1loopK}.}
  \label{fig::2}
\end{figure}

\subsection{Leading logarithmic evolution}
\label{sec::LL}

The $n$th-order double-logarithmic correction originates from
the planar $n$-loop annihilation ladder diagram
\cite{Gorshkov:1966ht}. The one-loop analysis in \cref{sec::one-loop}
can be  extended to any number of loops in a
straightforward manner. The electron propagator of each loop is
cancelled and the muon propagators effectively become scalar.
The corresponding effective theory diagram is given in
\cref{fig::2a}, which involves the soft on-shell muons
with loop momenta $\ell_i$, $i=1,\ldots , n$ and the propagators
$-2\pi i \delta(\ell_i^2-m_\mu^2)$, and the eikonal off-shell
transverse photons with propagators $g^{\perp}_{\mu\nu}/(2\ell_i \cdot \ell_j)$.
The double-logarithmic scaling of the $n$-loop diagrams
requires the hierarchy
\begin{align}
    \frac{m_\mu^2}{\sqrt{s}}< \ell^+_n<\ldots < \ell^+_1<\sqrt{s} \,, \qquad
    \frac{m_\mu^2}{\sqrt{s}}<\ell^-_1<\ldots< \ell^-_n < \sqrt{s}\,.
    \label{eqn:LL-ladder-lightcone-ordering}
\end{align}
These rules can be used for the  direct evaluation  and
resummation of the double-logarithmic terms. We, however,
reformulate the problem in the spirit of renormalization-group
flow, and introduce an auxiliary ultraviolet cutoff $\nu^+$ on
the plus-components of the light-cone momenta and an infrared
cutoff $\nu^-$ on the minus-components, with
$m_\mu^2/\sqrt{s}<\nu^\pm<\sqrt{s}$. Within these cutoffs the
effective theory description discussed in the previous section
is valid, and our main goal is to find the cutoff dependence of
the effective theory amplitude. Although it does not affect
the leading logarithmic approximation, it is convenient for a
further analysis to write the amplitude of the process in the
form
\begin{equation}
{\cal M}={\alpha(\sqrt{s})\over \alpha(m_\mu)}{\cal Z}_{e\mu}A(z)\,
{\cal M}_{\rm Born}+\ldots\,,
\label{eq::decomp}
\end{equation}
where
the factor ${\cal Z}_{e\mu}$ describes the collinear
renormalization of the external on-shell lines
\cite{Anastasiou:2020vkr}. In the leading single-logarithmic
approximation it is given by
\begin{equation}
{\cal Z}_{e\mu}=
\left( \frac{s}{m_e
m_\mu} \right)^{{\alpha\over \pi}\gamma^{(1)}_f}\,.
\label{eq::Zem}
\end{equation}
The form factor $A(z)$ is a function  of  the variable
\begin{equation}
z = \frac{\alpha L^2}{2\pi}\,,
\label{eq::zdef}
\end{equation}
and accommodates the double logarithmic terms to all orders
in $\alpha$. We only consider the real part of the form factor
since its imaginary part does not contribute to the
cross section through the NLL approximation. In the effective
theory the form factor  becomes a function of the cutoffs
$A(\xi,\eta)$, where we introduce the normalized logarithmic
variables
\begin{align}
    \label{eqn:xi-eta-LL}
    \xi=\ln(\nu^+\sqrt{s}/m^2_\mu)/L \,, \qquad
    \eta=-\ln(\nu^-/\sqrt{s})/L \,,
\end{align}
within $0 < \eta < \xi < 1$. The full-theory
amplitude is then given by $A(z)=A(1,1)$ corresponding to the
physical cutoffs $\nu^+=\sqrt{s}$, $\nu^-=m_\mu^2/\sqrt{s}$.
The double-logarithmic region is now restricted to
\begin{align}
    \label{eqn:LL-region-cutoffs}
    \ell^+_n<\ldots < \ell^+_1<\nu^+ \,, \qquad
    \nu^-<\ell^-_1<\ldots< \ell^-_n\,.
\end{align}
In this way, the second-order
derivative with respect to $\xi$ and $\eta$ captures the
ultraviolet and infrared singularities of the  $\ell_1$ integral,
and  maps the $n$-loop into the $(n-1)$-loop effective
theory diagram. In other words, the change of the all-order leading
logarithmic form factor with the variation of the cutoffs is
proportional to the form factor itself. This yields the
following evolution equation
\begin{equation}
{\partial^2 A(\xi,\eta)\over \partial\xi\partial\eta} \,
=z A (\xi,\eta)\,,
\label{eq::evoleqLL}
\end{equation}
with  boundary conditions
\begin{align}
    \label{eq::bcll1}
    A(\xi,0) = {} & 1\,, \\
    \label{eq::bcll2}
    \frac{\partial A(\xi,\eta)}{\partial \eta}
    \bigg\rvert_{\eta=\xi}={} &0 \,.
\end{align}
The first condition indicates that when the infrared cutoff
approaches the physical hard scale $\nu^-=\sqrt{s}$, the
effective-theory phase space shrinks to nothing and the
amplitude reduces to the Born result.  The second boundary
condition is purely kinematical. The soft muon on-shell
condition requires $\ell^-_i\ell^+_i>m_\mu^2$.
Thus, since $\ell^+_i<\nu^+$, the virtual momentum  region
$\nu^-<\ell^-_i<m_\mu^2/\nu^+$ does not include the pole of the
muon propagator and does not contribute to the double
logarithmic integral. Hence, for $\nu^-\le m_\mu^2/\nu^+$
corresponding to $\eta\ge \xi$ the amplitude does not depend on
the infrared cutoff. The solution of
\cref{eq::evoleqLL,eq::bcll1,eq::bcll2} reads
\begin{equation}
\ALL(\xi,\eta)=I_0\left(2(z\eta\xi)^{1/2}\right)
-{\eta\over \xi}I_2\left(2(z\eta\xi)^{1/2}\right)\,,
\label{eq::Ares}
\end{equation}
where $I_n(x)$ is the $n$th modified Bessel function.
This recovers the leading logarithmic result
\cite{Gorshkov:1966ht}
\begin{equation}
\ALL(z)={I_1(2z^{1/2})\over z^{1/2}} \,.
\label{eq::A11}
\end{equation}
Expanding this result for $z\to0$,
\begin{align}
    \ALL(z)=1+{z\over 2}+{z^2\over 12} + \mathcal{O}\left(z^3\right) \,,
\end{align}
corresponds to the fixed-order leading logarithmic behavior
at one (cf.~\cref{sec::one-loop}) and two loops. The asymptotic
behavior of the form factor at high energies, $z\to\infty$,
reads
\begin{align}
 \ALL(z) = \frac{e^{2\sqrt{z}}}{2\sqrt{\pi} z^{3/4}}
 \left[ 1 + \mathcal{O}\left(\frac1{z^{1/2}}\right)  \right] \,.
\end{align}

\subsection{Next-to-leading logarithms}
\label{sec::NLL}

\begin{figure}[t]
  \begin{center}
\begin{tabular}{c}
  \hspace*{0mm} \raisebox{61mm}{\small${p_1}$}
  \hspace*{32.5mm}
  \hspace*{0mm} \raisebox{61mm}{\small${p_3}$}
  \hspace*{-43.5mm}
  \raisebox{52.5mm}{\small$\ell_{n}$} 
  \hspace*{-6mm} \raisebox{37.6mm}{\small${\ell_{i+1}}$}
  \hspace*{-8.5mm} \raisebox{29.5mm}{\small${\ell}$}
  \hspace*{-4mm} \raisebox{21.1mm}{\small${\ell_{i}}$}
  \hspace*{-4.5mm} \raisebox{6.1mm}{\small${\ell_{1}}$}
  \hspace*{-9.5mm} \raisebox{-2mm}{\small${p_2}$}
    \hspace*{33mm}
  \hspace*{0mm} \raisebox{-2mm}{\small${p_4}$}
  \hspace*{-52mm}
  \includegraphics[width=5.6cm]{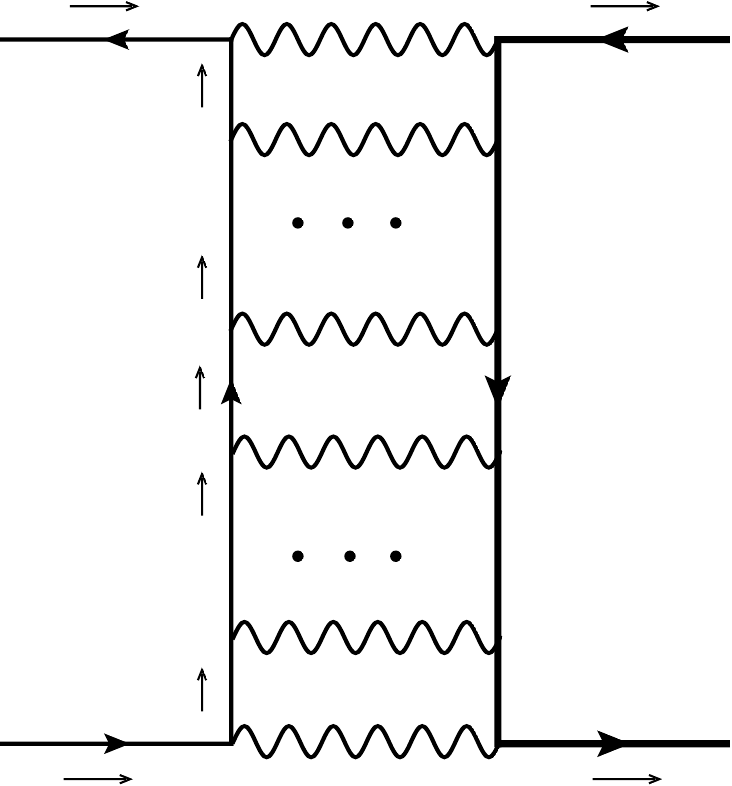}
\end{tabular}
\end{center}
  \caption{
  The  $(n+1)$-loop QED Feynman diagram determining the photonic
  NLL contribution to the evolution equation kernel. Here 
  $\ell_i$, $i=1,\ldots n$ are the LL loop momenta and $\ell$ is 
  the NLL loop momentum.}
  \label{fig::3}
\end{figure}

In  the next-to-leading logarithmic approximation the scale
factorization is rather straightforward.  Indeed, the collinear
logarithms factorize and are captured by an overall ${\cal
Z}_{e\mu}$, cf.~\cref{eq::decomp,eq::Zem}. Then, the leading
non-Sudakov logarithms specific to the forward kinematics
originate from a planar box with the external momentum transfer
{\it smaller} than the fermion mass. At the same time a soft
photon exchange outside the planar box, as in \cref{fig::2b},
develops a logarithmic scaling only when the corresponding loop
momentum components are {\it larger} than the fermion mass,
{\it i.e.} the two logarithmic momentum regions do not overlap.
Thus such an exchange either  produces pure Sudakov
logarithms, which cancel out in the total forward amplitude, or
contributes at the next-to-next-to-leading logarithmic order.
Hence, the next-to-leading logarithmic approximation is
determined by the sum of the diagrams in \cref{fig::2c}.
These are obtained from \cref{fig::2a} by inserting the single
logarithmic soft contribution to the one-loop (almost) on-shell
forward annihilation amplitude with the electron carrying the
leading logarithmic loop momenta $\ell_{i}$ and the positron
carrying the leading logarithmic loop momenta $-\ell_{i+1}$,
with $i=0,\ldots,n$, $\ell_0=p_2$, and $\ell_{n+1}=-p_1$ for
$(n+1)$-loop diagram. We have verified through the explicit
two-loop calculation that the sum of all other  diagrams indeed
contributes only to the prefactor \cref{eq::Zem},
confirming the above factorization picture. The relevant
one-loop soft single-logarithmic terms have been analyzed in
\cref{sec::one-loop}. For $m_e\ll m_\mu$, they include the
renormalization group running of $\alpha$ and the photonic
contribution of \cref{eq::slint}. We begin by considering
the photonic part; the corresponding QED Feynman diagram is
given in \cref{fig::3}. The integral over the loop
momentum $\ell$ results in the NLL contribution. It can be
obtained directly from \cref{eq::slint} by substituting
$Q^2\to 2\ell_{i}\cdot \ell_{i+1}$, which gives

\begin{equation}
-{\alpha L\over 2\pi}(\xi_{i}-\eta_{i+1})\,,
\label{eq::1loopphoton}
\end{equation}
as a correction factor to the
integrand of the $n$-loop LL diagram. In \cref{eq::1loopphoton}
we have introduced logarithmic variables
\begin{equation}
    \xi_i=\ln(\ell_i^+\sqrt{s}/\mEsq)/L \,, \qquad
    \eta_i=-\ln(\ell_i^-/\sqrt{s})/L\,,
    \label{eq::xiietaidef}
\end{equation}
for the integration over the light-cone components of the $i$th
virtual momentum, where $\xi_0=1$ and $\eta_{n+1}=0$.
\footnote{Strictly speaking, $\eta_{n+1}$ would develop
an imaginary part that we can disregard at NLL order.} 
Next, we consider the vacuum polarization correction to the 
photon propagator carrying the momentum $\ell_{i}-\ell_{i+1}$.
In this case, we obtain the correction factor
\begin{equation}
-{\alpha L\over 2\pi}{\beta_0\over 2}(\xi_{i}-\eta_{i+1})\,.
\label{eq::1loopbeta}
\end{equation}
Combining both results, the total NLL soft contribution
can be written as a sum over the effective theory
diagrams in~\cref{fig::2c}, 
\begin{equation}
 \sum_{i=0}^nK(\xi_{i}-\eta_{i+1})\,,
\label{eq::Ksum}
\end{equation}
where the soft kernel reads
\begin{equation}
K(\zeta)=
-{\alpha L\over 2\pi}\left({\beta_0\over 2}
+1\right) \zeta \,.
\label{eq::1loopK}
\end{equation}
With the cutoffs  \cref{eqn:LL-region-cutoffs}, the interval of
the remaining double-logarithmic integration is
\begin{align}
    0<\xi_n<\ldots <\xi_1<\xi\,, ~~~ 0<\eta_n<\ldots <\eta_1<\eta\,, ~~~ \eta_i<\xi_i\,.
\end{align}
Thanks to the linearity of $K(\zeta)$, we can re-arrange the sum in \cref{eq::Ksum}
\begin{align}
\label{eq::Ksum-rearranged}
 \sum_{i=0}^n K(\xi_{i}-\eta_{i+1})  = K(1) + \sum_{i=1}^n K(\xi_{i}-\eta_{i})\,.
\end{align}
Hence, the $i=1$ term of \cref{eq::Ksum-rearranged}
gives the NLL contribution to the evolution equation, which now 
takes the form
\begin{equation}
{\partial^2 A(\xi,\eta)\over\partial\xi\partial\eta} \,
=z\left(1+K(\xi-\eta)\right)A(\xi,\eta)\,.
\label{eq::evoleqNLLexp}
\end{equation}
The correction term in \cref{eq::evoleqNLLexp} is ${\cal
O}(\alpha^2)$ \textit{i.e.} it defines the evolution starting
from two loops. The one-loop contribution corresponds to $K(1)$, \textit{i.e.} $n=0$, in \cref{eq::Ksum-rearranged}. Since it does not depend on loop momenta, the same constant multiplies the LL amplitude at all orders. Therefore, its photonic part is accounted for by matching the  boundary condition  \cref{eq::bcll1} to the one-loop amplitude
\begin{equation}
A(\xi,0)=1 -{\alpha L\over
2\pi}\,.
\label{eq::bcnll}
\end{equation}
We note that compared to the r.h.s of \cref{eq:EM-1L-NLL-coeff},
$\gamma^{(1)}_f$ and  $\beta_0$ are absent in \cref{eq::bcnll}
due  to our definition of the form factor $A(\xi,\eta)$. Their
contribution to the amplitude arises from the expansion of the
factor $\alpha(\sqrt{s})/\alpha(m_\mu){\cal Z}_{e\mu}$ in
\cref{eq::decomp}.  The second  boundary condition
\cref{eq::bcll2} is purely kinematical and hence does not
change.

The electron mass effect can be accommodated in the evolution
equation as follows. \cref{eq::mecont} represents the
difference between the  LL contributions with the different
infrared scales. It merely reflects the reduction  of the
volume of the double logarithmic integration region due to the
electron mass and hence can be accounted for by adjusting the
infrared scale of the large logarithm from $m_\mu^2$ to
$e^{-\gamma^{(1)}_e}m_\mu^2$,  so that 
\begin{equation}
L\to L'= \ln\left(e^{\gamma^{(1)}_e}{s\over
m_\mu^2}\right)=L+\gamma^{(1)}_e\,,
\label{eq::Lprime}
\end{equation}
and the $\gamma^{(1)}_e$ term disappears from
\cref{eq:EM-1L-NLL-coeff}. Thus, the electron mass effect
results in a redefinition of the double logarithmic variable to
$z={\alpha\over 2 \pi}{L'}^2$ and can be omitted in the NLL
kernel \cref{eq::1loopK}. 
We have checked this by the explicit evaluation
of the two-loop planar box diagram in the OM and EM limits. 

The differential equation in \cref{eq::evoleqNLLexp} can
be solved perturbatively about the leading-logarithmic result.
The solution reads
\begin{align}
\ANLL(\xi,\eta)={}&\bigg(1-{\alpha L\over 2\pi}
\bigg) \ALL (\xi,\eta)
\nonumber \\
+&{}z\int_{0}^1{\rm d}\eta'\int_{\eta'}^1{\rm d}\xi'G(\xi-\xi',\eta-\eta')
K(\xi'-\eta') \ALL(\xi',\eta')\,,
\label{eq::eqintNLL}
\end{align}
where $G(\xi-\xi',\eta-\eta')$ is the Green's function of
\cref{eq::evoleqLL}
\begin{equation}
\left({\partial^2 \over \partial\xi\partial\eta}
-z\right)G(\xi-\xi',\eta-\eta')=\delta(\xi-\xi')\delta(\eta-\eta')\,.
\label{eq::GF}
\end{equation}
We are interested in the solution for $\xi=\eta=1$ and do not
need the explicit form of the Green's function since for
$\xi'<\xi,~\eta'<\eta$ it is given by a solution of the
homogeneous equation. With the given boundary conditions, for
$\xi',~\eta'<1$ we have
\begin{equation}
G(1-\xi',1-\eta')=\ALL(1-\eta',1-\xi')\,.
\label{eq::GFsol}
\end{equation}
Thus, the next-to-leading logarithmic  result for the
form factor is given by
\begin{align}
\ANLL(z) =
\bigg[1-{\alpha L\over 2\pi}\bigg] \ALL(z)
-{\alpha L\over 2\pi}\left({\beta_0\over 2
}+1\right) k(z)\,,
\label{eq::MNLL}
\end{align}
where
\begin{align}
k(z) = z\int_{0}^1{\rm d}\eta'\int_{\eta'}^1{\rm d}\xi'
(\xi'-\eta')\ALL(1-\eta',1-\xi')\ALL(\xi',\eta')
\,,
\label{eq::kz}
\end{align}
and $\ALL(z)$ is given in \cref{eq::A11}. The first ten terms of
the  Taylor expansion $k(z)=\sum_{n=1}^\infty k_nz^{n}$ are  given
in \cref{tab::1}. The asymptotic behavior of $k(z)$ for
$z\to\infty$ is derived in \cref{app:asyexp-k} and reads
\begin{align}
    k(z) = \frac{e^{2\sqrt{z}}}{ 4\sqrt{z}}
    \left[ 1 + \ORD{\frac1{z^{1/4}}} \right] \,.
\end{align}
We can perform a nontrivial check of our perturbative solution
of \cref{eq::evoleqNLLexp} by taking a cutoff-independent
perturbation due  to the variation  $z\to z+\delta z$. The
corresponding correction to the form factor  can be obtained
from \cref{eq::eqintNLL} with $K(\zeta)=\delta z / z $. On the
other hand it can be obtained directly by the Taylor expansion
of the leading logarithmic result in $\delta z$. This implies a
nonobvious nonlinear integral  relation between the modified
Bessel functions
\begin{align}
\label{eq::Bessrel}
\int_{0}^1{\rm d}\eta'\int_{\eta'}^1{\rm d}\xi'
\ALL(1-\eta',1-\xi')  \ALL(\xi',\eta')
={d \ALL(z)\over d z}
={I_2(2z^{1/2})\over z}\,,
\end{align}
which can be easily  verified. For $\delta z$ corresponding to
the renormalization scheme variation of the coupling constant,
\cref{eq::Bessrel} ensures the  scheme invariance of the
result.

We have verified by the explicit two-loop calculation that the
NLL corrections in \cref{eq::MNLL} stem from the two-loop
planar box diagram and the  vacuum polarization correction to
the one-loop planar box diagram, while the rest contributes to
the factor \linebreak 
$\alpha(\sqrt{s})/\alpha(m_\mu){\cal Z}_{e\mu}$ in
\cref{eq::decomp}, in full agreement with the factorization of
the collinear and cancellation of the soft logarithms discussed
in the beginning of the section.

\begin{table}[t]
  \begin{center}
    \begin{tabular}{|c|c|c|c|c|c|c|c|c|c|c|}
     \hline
      $n$ & $1$ & $2$ & $3$ & $4$ & $5$ & $6$ & $7$  &$8$  &$9$  &$10$ \\
      \hline
      & &  & &  &  &  &  & & & \\[-3mm]
      $6(n!)^2 \, k_n$ & 1 & ${6\over 5}$ & ${87\over  70}$ & ${26\over 21}$&
      ${281\over 231}$ & ${170\over 143 }$ &
      ${9949\over 8580}$ & ${13742\over  12155}$&
      ${254649\over 230945}$&
      ${31630\over  29393}$
      \\[1.5mm]
      \hline
    \end{tabular}
   \end{center}
    \caption{\label{tab::1} The   normalized coefficients of
    the Taylor series for the function $k(z)$   up to  $n=10$.}
\end{table}

By expanding the full amplitude in \cref{eq::decomp} in the
coupling constant with $A(z)$ given by \cref{eq::MNLL} we get
the two-loop NLL contribution to the amplitude
\begin{align}
{\cal M}^{(2)}
=\left[{L^4\over 12}
+\left({\gamma^{(1)}_e\over 3}+{5\over6}
+{4\over 9}n_l\right)L^3\right]{\cal M}_{\rm Born}+\ldots\,,
\label{eq::MNLL2l}
\end{align}
where we keep only the terms that contribute to the cross
section. This calculation  provides  the ${\cal O}(\alpha^2)$
virtual corrections to the cross  section of $e^+ e^- \to
\mu^+\mu^-$ annihilation in forward kinematics through the
next-to-leading logarithms for arbitrary electron-to-muon mass
ratio. Writing $d\sigma=\sum_{n=0}^\infty({\alpha\over
2\pi})^{n}d\sigma^{(n)}$ for the one and two-loop terms we get
\begin{align}
\label{eqn:sigmaNLO}
{d\sigma^{(1)}\over d\sigma_{\rm Born}} = {}&
L^2+\left(2\gamma^{(1)}_e+4+{4\over 3}n_l\right) L\,, \\
\label{eqn:sigmaNNLO}
{d\sigma^{(2)}\over d\sigma_{\rm Born}} = {} &
{5\over 12}L^4
+\left(\frac{5}{3}\gamma^{(1)}_e +{11\over 3}  +{14\over 9}n_l\right)L^3\,,
\end{align}
where the coupling constant in the Born cross section is
renormalized at $m_\mu$. This result is valid for an 
{\it arbitrary} electron-to-muon mass ratio.

While \cref{eqn:sigmaNLO,eqn:sigmaNNLO} are finite, for unequal
masses the virtual corrections to the cross section become
infrared divergent starting from the next-to-next-to-leading
logarithmic approximation. To obtain the physical and finite
contribution to the cross section from the forward region, the
real emission of soft photons with characteristic energy
$\varepsilon_\gamma\lsim m_\mu^2/\sqrt{s}$ should be added, which
factors out with respect to the virtual corrections discussed
above. In the equal-mass case, the cross section is infrared
finite beyond the logarithmic approximation since the soft
emission vanishes identically along with the momentum transfer.
\cref{eqn:sigmaNNLO} agrees  with the result of the explicit
analytical evaluation of the two-loop terms surviving the
high-energy limit for both OM and EM cases described in detail
in \cref{sec::2LFO}.

\subsection{Regularization and scheme dependence}
\label{sec::NLLad} 
Though the NLL contribution to the amplitude is infrared
finite, the parameters of the evolution equation depend in
general on the regularization and subtraction scheme for
infrared divergences.  If a physical infrared regulator is
used, {\it e.g.} the auxiliary photon mass as it is done in
\cref{sec::one-loop}, soft emissions with
energy smaller than $m_\gamma$ vanish, and virtual
corrections can be unambiguously  separated from the real
radiation. Then, the NLL coefficients of
\cref{eqn:sigmaNLO,eqn:sigmaNNLO} do not depend on
the infrared subtraction.  

In contrast, dimensional regularization does not eliminate
the real radiation of vanishing energy and one has to
incorporate the soft radiation to get a scheme-independent
result. As a consequence, even the finite NLL coefficients of
the virtual corrections become scheme dependent.
The infrared renormalization prescription for the dimensionally
regulated amplitude which recovers the result with a physical
regulator is given in \cref{eqn:definition-IR-subtraction}.
In dimensional regularization, the infrared singular terms
in the virtual corrections factorize and exponentiate in the
same way as for the real soft emission, and the subtraction can
be interpreted as their cancellation against the real emission
of photons with vanishing energy. 
Due to the exponentiation of the singular
terms the all-order equivalence to the physical regularization
can be established through the one-loop calculation. We have
verified this equivalence by the explicit evaluation
of the cross section through ${\cal O}(\alpha^2)$.

\subsection{Refined form of the evolution equation}
\label{sec::NLL-beyond}

Following the spirit of the effective theory renormalization
group we can refine the NLL analysis of \cref{sec::NLL}, taking
into account  the running of the coupling constant in the
evolution equation \textit{without} the expansion in $\alpha$.
The corresponding leading renormalization group running
accounts for the terms of the form $\alpha^nL^n$ which are
formally beyond the next-to-leading approximation limited to
$\alpha^nL^{2n-1}$ terms. At the same time, such a refined form
emphasizes the factorization structure and sets the stage for
the higher-order analysis. In
the $n$-loop leading logarithmic contribution the $i$th photon
propagator is off-shell by $\ell_i^+\ell_{i+1}^-\propto
e^{{L}(\xi_{i}-\eta_{i+1})}m^2_\mu$ which gives the vacuum
polarization corrections  of the form
\begin{equation}
{\alpha(e^{{L\over 2}(\xi_{i}-\eta_{i+1})}m_\mu)\over\alpha}
=1-{\alpha L\over 4\pi}\beta_0(\xi_{i}-\eta_{i+1})+\ldots\,.
\label{eq::bt0vpf}
\end{equation}
Due to the momentum ordering in the leading logarithmic
contribution, the derivatives with respect to the cutoffs act
on the components of the same loop momenta. \cref{eq::bt0vpf},
however, has a complex dependence on neighboring loop momenta
which only separates through ${\cal O}(\alpha)$. To accommodate
this dependence, we first note that the differential equation
\cref{eq::evoleqLL} is equivalent to the integral equation
\begin{equation}
A(\xi,\eta)=z\int_{0}^\eta{\rm d}\eta'\int_{\eta'}^\xi{\rm d}\xi'\,
A(\xi',\eta')+ A(0,0)\,.
\label{eq::inteqLL}
\end{equation}
Substituting it into the right hand side of
\cref{eq::evoleqLL}, we get
\begin{equation}
{\partial^2 A(\xi,\eta)\over \partial\xi\partial\eta}
=z^2\int_{0}^\eta{\rm d}\eta'\int_{\eta'}^\xi{\rm d}\xi'\,
A(\xi',\eta')+ zA(0,0)\,.
\label{eq::intevoleqLL}
\end{equation}
Now  $\xi$ and $\eta'$ correspond to the cutoff on the
neighboring loop momenta components, and we can account for the
all-order running of the coupling by inserting the vacuum
polarization contribution \cref{eq::bt0vpf} with the
corresponding arguments into the right hand side of
\cref{eq::intevoleqLL}. The evolution equation then
becomes
\begin{align}
{\partial^2 A(\xi,\eta)\over \partial\xi\partial\eta}
={} & z\left(1-{\alpha L\over 2\pi}(\xi-\eta)
\right)A(\xi,\eta)
\nonumber\\
+{} & z^2
\int_{0}^\eta{\rm d}\eta'\int_{\eta'}^\xi{\rm d}\xi'\,
\Delta(\xi-\eta')A(\xi',\eta')
+z\Delta(\xi)A(0,0)\,.
\label{eq::evoleqNLL}
\end{align}
where
\begin{equation}
\Delta(\zeta)={\alpha(e^{{L\over 2}\zeta}m_\mu)
\over \alpha}-1={1\over 1+{\alpha L\over 4\pi}\beta_0\zeta}-1\,.
\label{eq::del}
\end{equation}
The  series in  $\alpha$ for  the right hand side of
\cref{eq::evoleqNLL}  differs from  \cref{eq::evoleqNLLexp} at
${\cal O}(\alpha^2)$. Hence, this form of the evolution
equation requires matching of the $\beta_0$ terms through two
loops, which can be done by adding the corresponding  boundary
term given by the convolution of the leading order solution
with the kernel difference. A rather complex
integro-differential equation in \cref{eq::evoleqNLL} can
presumably only be solved numerically, which is beyond the
scope of this paper.

\section{Two-loop corrections in  the high-energy forward limit}
\label{sec::2LFO}
In this section we provide the details of the calculation of
the two-loop forward annihilation amplitude in the high-energy
limit \cite{Delto:2025epy}. Two-loop QED corrections to
four-fermion scattering have already been analyzed for the
massless-massive case in~\cite{Mastrolia:2017pfy,Bonciani:2021okt}, 
for the equal-mass case in~\cite{Delto:2023kqv}, and in the limit of 
small electron mass in~\cite{Gerasimov:2025pqf}. These results,
however, are limited  to finite scattering angles and can not
be extrapolated to the forward region.

\subsection{Form-factor decomposition}

We adopt the ’t~Hooft--Veltman scheme (tHV) \cite{tHooft:1972tcz}, where internal degrees of freedom are analytically continued to $d=4-2\epsilon$ dimensions, whereas external states are kept four-dimensional. Following the approach of Refs.~\cite{Peraro:2019cjj,Peraro:2020sfm}, we decompose the forward-scattering amplitude for the process in \cref{eqn:def-process} into two relevant tensor structures and their respective form factors\footnote{In comparison to Ref.~\cite{Delto:2025epy}, $\mathcal{F}_1=\widetilde{\mathcal{F}}$ and $\mathcal{F}_2=\mathcal{F}$.}
\begin{align}
    \label{eqn:amp-FF-decomposition}
    i \mathcal{M} = -\frac{i e^2}{s} \sum_{i=1}^{2} T_i \mathcal{F}_i = -\frac{i e^2}{s} \left[ \frac{m_e}{m_\mu} \left( \overline{V}_e U_e \right) \left( \overline{U}_\mu  \, V_\mu \right) \mathcal{F}_1 + \left( \overline{V}_e \gamma^{\rho} U_e \right) \left( \overline{U}_\mu \gamma_{\rho}   \, V_\mu \right) \mathcal{F}_2  \right] \,.
\end{align}
We note that the particular choice of prefactors in the above equation ensures that the form factors are dimensionless and that they scale as $\mathcal{O}(1)$ in the high-energy limit for both the EM and the OM case. By construction, $\mathcal{F}_1$ is zero at tree-level and free of leading-logarithmic contributions at one loop and beyond.

Additional tensor structures that would generally arise in massive four-fermion scattering amplitudes are either linearly dependent on $T_{1,2}$ in the forward limit or not required to obtain the tHV result~\cite{Peraro:2019cjj,Peraro:2020sfm}. Following the standard approach, we compute the form factors in \cref{eqn:amp-FF-decomposition}
by defining a set of projection operators as combinations of \textit{dual tensors} $T_k^\dagger$
\begin{align}
    \label{eqn:def-projectors}
    \mathcal{P}_i = \left[ M^{-1} \right]_{ik} T_k^{\dagger} \,, ~~ M_{ij} = T_i^{\dagger} * T_j  ~~\Rightarrow ~~ \mathcal{P}_i *  (i\mathcal{M}) =  \frac{-ie^2}{s} \mathcal{F}_i \,,
\end{align}
where ``$*$'' denotes summation over the spins of external fermions. In terms of lepton velocities and our choice of tensor decomposition, the matrix reads
\begin{align}
\label{eqn:MMij-beta}
M ={} & 4 s^2
\begin{pmatrix}
\frac{\left(1-\be^2\right)\be^2\bE^2}{\left(1-\bE^2\right)} & -(1-\be^2)\be\bE\\
-(1-\be^2)\be\bE & 3-\be^2-\bE^2+\be^2\bE^2-2\epsilon
\end{pmatrix}
\,.
\end{align}

\subsection{Technical aspects}

We use a standard tool-chain, generating the Feynman diagrams with
\texttt{QGRAF}~\cite{Nogueira:1991ex},
inserting Feynman rules, applying projectors, and mapping onto scalar Feynman integrals
using the computer algebra system \texttt{FORM}~\cite{Vermaseren:2000nd,Kuipers:2012rf,Kuipers:2013pba,Ruijl:2017dtg}.
We employ the public tool \texttt{Reduze2}~\cite{vonManteuffel:2012np} to find mappings onto
integral topologies and to expose symmetries among integrals. In order to simplify the computation early on,
we then take the strict forward scattering limit both on the scalar integrals and on
their coefficients before IBP reduction. We are only interested in the leading behavior and parameterize the forward limit
using \cref{eqn:p1234-FS-COMF}.\footnote{We note that an extension to higher-order contributions in $(1-\cos\theta_{13})$ is a conceptually straightforward exercise.}
In particular, the forward limit also implies that outgoing momenta are not linearly independent, but linear combinations of the incoming ones
\begin{align}
p_{3,4} = \frac{1}{2} \bigg[ (1\pm \betafrac) p_1 + (1\mp \betafrac) p_2 \bigg] \,,
\end{align}
where $\betafrac=\bE/\be$. In this parametrization propagators become linearly dependent. We devise a simple partial-fraction strategy implemented in \texttt{Mathematica}
to map resulting loop integrals onto topologies that depend only on $p_{1,2}$,
and we use \texttt{Feynson} \cite{Maheria:2022dsq} to expose
additional symmetries that arise among integrals in forward-scattering kinematics after the partial-fraction decomposition.
Employing \texttt{Kira}~\cite{Maierhofer:2017gsa,Maierhofer:2018gpa,Klappert:2020nbg}
to solve the required integration-by-parts (IBP) relations \cite{Tkachov:1981wb,Chetyrkin:1981qh}
we express the one- and two-loop amplitudes through 9 and 191 master integrals, respectively.
This task is achieved following Laporta's algorithm~\cite{Laporta:2000dsw},
improved by finite field techniques~\cite{vonManteuffel:2014ixa,Peraro:2016wsq}.

With the help of \texttt{LiteRed}~\cite{Lee:2013mka} and the aforementioned IBP relations we then derive systems of first-order partial differential
equations for master integrals~\cite{Kotikov:1990kg,Kotikov:1991hm,Kotikov:1991pm,Remiddi:1997ny,Gehrmann:1999as}
as functions of $s$, $\be$, and $\bE$. We construct a canonical basis~\cite{Henn:2013pwa}
of master integrals, in which differential equations assume a particularly simple, {\it i.e.} $\epsilon$-factorized form.
They read
\begin{align}
    \mathrm{d} J^{(l)} = \epsilon \, \mathrm{d} A^{(l)} J^{(l)}\,,~~ l=1,2\,,
\end{align}
where the matrices $\mathrm{d} A^{(l)}$ consist of rational
numbers and one-forms $\mathrm{dlog}(\omega)$. At one loop, the integration kernels $\omega$ are drawn from the set
\begin{align}
\Omega^{1L} = \left\{ \be , \bE , 1 \pm \be , 1 \pm \bE \right\}\,,
\end{align}
and differential equations can be solved in terms of Harmonic Polylogarithms~\cite{Remiddi:1999ew}.
At two loops we find rational and algebraic integration kernels $\Omega^{2L} = \Omega^{2L,\text{rat}} \cup \Omega^{2L,\text{alg}}$, the latter containing square roots
\begin{align}
    r_{1,\dots,5}^2 =\{ 1-\be^2,1-\bE^2, 1-\be^2+\bE^2,1+\be^2-\bE^2,\be^2-\bE^2 \} \,.
\end{align}
While master integrals in the form described above are a  good starting point for numerical evaluation in general,
in the context of this work we are interested in extracting their high-energy behavior analytically.
For both mass configurations, this limit corresponds to regular-singular points of the differential equations.
Given their particular form, in order to solve the differential equations in terms of power-log series expansion, we require blow-ups,
which ensure that at most two singular lines intersect and that each of these intersections is non-tangential; we introduce the parametrizations
\begin{alignat}{3}
   &\text{EM:}~ &&\big\{ \be = 1 - 2 (1-\varBUeE^2) \varBUE^2 \,, ~~ &&\bE = 1-2 \varBUE^2  \big\} \,, \label{eqn:para-blowup-EM}\\
   &\text{OM:}~ &&\big\{ \be = 1 - 2 \varBUeE^2 \varBUE^4 \,, ~~ &&\bE = 1-2 \varBUE^2  \big\}  \,. \label{eqn:para-blowup-OM}
\end{alignat}
The high-energy limit corresponds to $\varBUE\to0$, with $\varBUeE\to0$ in the EM case and arbitrary $\varBUeE=\me \sqrt{s}/\mEsq$ in the OM case, on which the master integrals depend only logarithmically at leading power. We expand and solve the differential equations around $\varBUE=\varBUeE=0$ in both cases, resulting in solutions of the form
\begin{align}
    \label{eqn:J-power-log-exp}
  J^{(l)}   = \sum_{p=0}^{4} \epsilon^p
    \sum_{\alpha,\beta=0}^{N} \sum_{\gamma,\delta=0}^{p} \mathcal{C}^{(l)}_{p,\alpha,\beta,\gamma,\delta} \, \varBUE^\alpha \varBUeE^\beta \ln^\gamma(\varBUE) \ln^\delta(\varBUeE) \,,
\end{align}
where, depending on the loop order and mass configuration, we work to powers of $N=60,\dots,90$. We note that $\ln(\varBUeE)$ terms are absent in the case of equal masses as expected.

Almost all coefficients in \cref{eqn:J-power-log-exp} follow from
the differential equations themselves. We fix the remaining boundary constants $\mathcal{C}^{(l)}_{p,0,0,0,0}$
at $\varBUE=\varBUeE=1/10$ using numerical values for master integrals with around 150 digits
obtained with \texttt{AMFlow}~\cite{Liu:2022chg}. The high precision of this evaluation, together with the extensive series expansion up to high $N$, allows us to express
constants in terms of a few transcendental numbers using the PSLQ algorithm \cite{Bailey:1992TR,Bailey:1999nv}. We note that the leading-power evaluations of the various form factors only require the first few orders, $N \approx 5$. 

\subsection{UV and IR renormalization}

Having computed all required master integrals in both kinematic regimes, we proceed by removing
ultraviolet (UV) divergences from the form factors through multiplicative renormalization
of the bare fields, the bare coupling and the bare fermion mass
\begin{align}
    \label{eqn:FF-ren-def}
    \mathcal{F}_{i,\mathrm{ren}} = Z_{2,e}^{\text{OS}} \, Z_{2,\mu}^{\text{OS}} \, \mathcal{F}_{i,\mathrm{b}}\!\left[m_{\mathrm{b},e}(\me),m_{\mathrm{b},\mu}(\mE),\alpha_{\mathrm{b}}(\alpha)\right]\,,
\end{align}
where $Z_{2,e/\mu}^{\text{OS}}$ are the on-shell wave function renormalization factors for leptons, see~\cref{app:renconst}. We renormalize the QED coupling constant in the $\overline{\text{MS}}$ scheme using
\begin{align}
    \alpha_{\mathrm{b}}\left(\overline{\alpha}\right) ={} & Z_{\alpha}^{\overline{\text{MS}}} \, S_{\epsilon}^{-1} \, \mu^{2\epsilon} \overline{\alpha} \,,
\end{align}
 where $Z_{\alpha}^{\overline{\text{MS}}}$ is given in~\cref{eq:Zalpha}, $S_{\epsilon}= (4\pi e^{-\gamma_E})^\epsilon$ and $\mu$ is the mass scale of dimensional regularization. We call $C_{\overline{\alpha}\alpha}$ the constant
 required to transition to the on-shell coupling $\alpha \approx 1/137$ through
\begin{align}
    \overline{\alpha}\left(\alpha\right) ={} & C_{\overline{\alpha}\alpha} \alpha \,.
\end{align}
We note that the remaining $\mu$-dependence in the on-shell renormalized form factors in \cref{eqn:FF-ren-def} stems from dimensionally regulated IR poles. Through two loops, the on-shell mass renormalization factors are flavor-diagonal
\begin{align}
    m_{\mathrm{b}}(m) ={} & Z_{m}^{\text{OS}} m \,, ~~~ m\in\{\me,\mE\} \,.
\end{align}
We expand the renormalization constants in terms of the $\overline{\text{MS}}$ coupling
\begin{align}
    \label{eqn:renconst-exp-general}
    \mathcal{R} = 1 + \sum_{i=1}^{2} \left( \frac{\overline{\alpha}}{2\pi} \right)^i \delta \mathcal{R}^{(i)} \,, \mathcal{R}\in \left\{ Z_{2,e}^{\text{OS}} , Z_{2,\mu}^{\text{OS}}, Z_{\me}^{\text{OS}} , Z_{\mE}^{\text{OS}}, Z_{\alpha}^{\overline{\text{MS}}} \right\} \,,
\end{align}
and the coupling conversion factor in terms of the on-shell coupling
\begin{align}
    \label{eqn:coupling-conversion-exp}
    C_{\overline{\alpha}\alpha} = 1 + \sum_{i=1}^{2} \left( \frac{\alpha}{2\pi} \right)^i \delta C_{\overline{\alpha}\alpha}^{(i)} \,.
\end{align}
The coefficients in \cref{eqn:renconst-exp-general,eqn:coupling-conversion-exp} are given in \cref{app:renconst}.

It is convenient to implement the mass renormalization by expanding the fermion propagators in the bare amplitude about the renormalized mass before IBP reduction. In our case fermion propagators start appearing at one loop, so we only need to expand to first order
\begin{align}
    \label{eqn:expansion_massive_fermionprop}
    \frac{\widehat{p}+m_\text{b}}{p^2-m_\text{b}^2}
    =
    \frac{\widehat{p}+m}{p^2-m^2} + \left(\frac{\overline{\alpha}}{2\pi}\right) \,  m \, \delta Z_{m}^{\text{OS},(1)}  \,
    \frac{ 2 m \widehat{p} + (p^2 + m^2) }{\left[p^2-m^2\right]^2} + \mathcal{O}\left( \overline{\alpha}^2 \right)\,.
\end{align}
Applying this expansion to the one-loop form factors, we write
\begin{align}
     \mathcal{F}_{i,\mathrm{b}}^{(1)} =  \mathcal{F}_{i,\mathrm{b}}^{(1),0} + \delta Z_{\me}^{\text{OS},(1)} \mathcal{F}_{i,\mathrm{b}}^{(1),\me} +  \delta Z_{\mE}^{\text{OS},(1)}  \mathcal{F}_{i,\mathrm{b}}^{(1),\mE}\,,
\end{align}
where all three form factors on the r.h.s. depend only on renormalized masses, and the last two contain squared propagators. 
Combining the relations above and expanding renormalized form factors in \cref{eqn:FF-ren-def} in powers of the on-shell coupling $\alpha$ yields
\begin{align}
    \mathcal{F}_{i,\mathrm{ren}} = \delta_{i2} +  \sum_{l=1}^{2} \left(\frac{\alpha}{2\pi} \right)^l \mathcal{F}_{i,\mathrm{ren}}^{(l)} \,.
    \label{eqn:Firen-pert-exp}
\end{align}
The coefficients $\mathcal{F}_{i,\mathrm{ren}}^{(l)}$ can be obtained from their bare counterparts using \cref{eqn:FFren-from-bare-0L,eqn:FFren-from-bare-1L,eqn:FFren-from-bare-2L}.

After UV renormalization we are left with one-loop exact infrared (IR) poles~\cite{Yennie:1961ad}
\begin{align}
    \label{eqn:definition-IR-subtraction}
    \mathcal{F}_{i,\mathrm{fin}} = \exp \left[ - \frac{\alpha}{2\pi} \frac{z_{\text{IR}}}{\epsilon}  \right] \mathcal{F}_{i,\mathrm{ren}}\,,
\end{align}
where $\mathcal{F}_{i,\mathrm{fin}}$ are finite in four dimensions
and $z_{\textrm{IR}}$ denotes the  anomalous dimension
that governs the  soft singularities of the amplitude to
all-orders through exponentiation~\cite{Becher:2009kw,Ferroglia:2009ep}. Again, we expand
\begin{align}
    \mathcal{F}_{i,\mathrm{fin}} = \delta_{i2} +  \sum_{l=1}^{2} \left(\frac{\alpha}{2\pi} \right)^l \mathcal{F}_{i,\mathrm{fin}}^{(l)} \,.
    \label{eqn:Fifin-pert-exp}
\end{align}
Taking the two kinematic limits, we obtain
\begin{align}
\label{eqn:ZIR-EM}
 z_{\text{IR}}^{\text{EM}} = {} & -2\pi i \,, \\
 \label{eqn:ZIR-OM}
 z_{\text{IR}}^{\text{OM}} = {} & -2\pi i -2 + \LEe \,.
\end{align}
Note that in the equal mass limit in \cref{eqn:ZIR-EM}, IR poles of virtual corrections turn into a pure phase. Hence, the corresponding cross section is IR finite and can be defined without the need to include real-emission corrections as would be required by the Bloch--Nordsieck theorem~\cite{Bloch:1937pw}. The reason is that, as already mentioned, the momentum exchanged between incoming and outgoing leptons of the same mass vanishes in the forward limit, which means that no charges are accelerated. We now turn to presenting results in the following section.

\subsection{Results and checks}
In the case of ordered masses, we further decompose one- and two-loop coefficients of the finite form factors in \cref{eqn:Fifin-pert-exp} as
\begin{align}
    \label{eqn:FF-ren-res-exp-OM}
    \mathcal{F}_{i,\mathrm{fin},\mathrm{OM}}^{(l)}= \sum_{\substack{a,b=0 \\ a+b \le l}}^{l} (\ne)^a (\nE)^b \left[  \mathcal{F}^{(l,a,b,\text{re})}_{i,\mathrm{fin},\mathrm{OM}} + i \pi \mathcal{F}^{(l,a,b,\text{im})}_{i,\mathrm{fin},\mathrm{OM}} \right] \,, ~~ l=1,2\,,
\end{align}
where $n_e$($n_\mu$) denotes the number of massive electrons (muons). In the equal masses case, the lepton contributions become the same  and we write
\begin{align}
    \label{eqn:FF-ren-res-exp-EM}
    \mathcal{F}_{i,\mathrm{fin},\mathrm{EM}}^{(l)} = \sum_{a=0}^{l} (\ne+\nE)^a \left[  \mathcal{F}^{(l,a,\text{re})}_{i,\mathrm{fin},\mathrm{EM}} + i \pi \mathcal{F}^{(l,a,\text{im})}_{i,\mathrm{fin},\mathrm{EM}} \right] \,, ~~ l=1,2\,.
\end{align}
The coefficients in \cref{eqn:FF-ren-res-exp-OM,eqn:FF-ren-res-exp-EM} constitute one of the main results of our work. 
We obtain them by combining the various ingredients discussed above, switching back to physical parameters $\mesq,\mEsq,s$ by inverting \cref{eqn:para-blowup-EM,eqn:para-blowup-OM}, and neglecting power corrections in the high-energy limit in final expressions. 
We note that the cancellation of UV and IR poles constitutes a strong check of the results, which are displayed in \cref{sec:res-OM,sec:res-EM}. Furthermore, we note that we find full agreement with the evolution equation results in \cref{eqn:sigmaNLO,eqn:sigmaNNLO}. 
For this comparison, we evaluate the results there in the EM or OM case, we set $\ne=\nE=\nl/2$, and we relate form factors to virtual contributions to the cross section using
\begin{align}
    \frac{\mathrm{d}\sigma^{(1)}}{\mathrm{d}\sigma_{\mathrm{Born}}} ={} \frac{2 \mathcal{K}_{0,1}}{\mathcal{K}_{0,0}} \,, ~~~~ \frac{\mathrm{d}\sigma^{(2)}}{\mathrm{d}\sigma_{\mathrm{Born}}} = \frac{2 \mathcal{K}_{0,2} + \mathcal{K}_{1,1}}{\mathcal{K}_{0,0}} \,,
\end{align}
where
\begin{align}
 \mathcal{K}^{\text{kin}}_{l,r} = \Re \left\{ \sum_{i,j=1}^{2} \left[ \mathcal{F}_{i,\text{fin},\text{kin}}^{(l)} \right]^* M^{\text{kin}}_{ij} \, \mathcal{F}_{j,\text{fin},\text{kin}}^{(r)} \right\}\,, ~~ \text{kin}\in \{\text{EM},\text{OM}\} \,.
\end{align}
We note that the matrix $M$ becomes diagonal in the high-energy limit for both mass configurations. Up to power corrections, we obtain
\begin{align}
M^{\text{OM}} = s^2\,
\begin{pmatrix}
0 & 0\\
0 & 8
\end{pmatrix}
\,,~~
M^{\text{EM}} =
s^2\, \begin{pmatrix}
4 & 0\\
0 & 8
\end{pmatrix}
\,.
\end{align}
Since $\mathcal{F}_1$ is zero at tree-level and free of leading logarithms, we can conclude from the diagonality of $M$ that only $\mathcal{F}_2$ contributes to the cross section through NLL, in full agreement with the discussion in \cref{sec::one-loop}.

\subsubsection{Ordered masses case}
\label{sec:res-OM}
We begin with the physically relevant case of ordered masses $\me \ll \mE \ll \sqrt{s}$. At a scale $\mu^2=\me \mE$ the one-loop coefficients in \cref{eqn:FF-ren-res-exp-OM} read
\begingroup
\allowdisplaybreaks
\begin{align}
\label{eqn:res-FF1-OM-1L}
&\mathcal{F}_{1,\text{fin},\text{OM}}^{(1,0,0,\text{re})} ={} - L \LEe - \bigg[ \frac1{2} \LEe^2 - 2 \LEe  \bigg] \,,
\quad \mathcal{F}_{1,\text{fin},\text{OM}}^{(1,0,0,\text{im})} ={} \LEe \,,  \\
& \mathcal{F}_{1,\text{fin},\text{OM}}^{(1,1,0,\text{re})} = \mathcal{F}_{1,\text{fin},\text{OM}}^{(1,1,0,\text{im})} = \mathcal{F}_{1,\text{fin},\text{OM}}^{(1,0,1,\text{re})} = \mathcal{F}_{1,\text{fin},\text{OM}}^{(1,0,1,\text{im})} = 0 \,, \nonumber
\end{align}
\endgroup
for the first form-factor and
\begingroup
\allowdisplaybreaks
\begin{alignat}{4}
    \label{eqn:res-FF2-OM-1L}
    & \mathcal{F}_{2,\text{fin},\text{OM}}^{(1,0,0,\text{re})} ={} && \frac{1}{2}L^2 + 2 L + \bigg[ \frac{3}{2} \LEe -5  + \frac{7\pi^2}{6}    \bigg] \,, \qquad
    && \mathcal{F}_{2,\text{fin},\text{OM}}^{(1,0,0,\text{im})} ={} && L + \bigg[ \LEe -2 \bigg] \,, \nonumber \\
    & \mathcal{F}_{2,\text{fin},\text{OM}}^{(1,1,0,\text{re})} ={}&& \frac{2}{3} L + \bigg[ \frac{2}{3} \LEe - \frac{10}{9} \bigg] \,, \qquad
    && \mathcal{F}_{2,\text{fin},\text{OM}}^{(1,1,0,\text{im})} ={} && -\frac{2}{3} \,, \\
    & \mathcal{F}_{2,\text{fin},\text{OM}}^{(1,0,1,\text{re})} ={}&& \frac{2}{3} L  - \frac{10}{9}  \,, \qquad
    && \mathcal{F}_{2,\text{fin},\text{OM}}^{(1,0,1,\text{im})} ={} && -\frac{2}{3} \,, \nonumber
\end{alignat}
\endgroup
for the second. At two loops, the non-zero contributions to the first form factor read
\begingroup
\allowdisplaybreaks
\begin{align}
\label{eqn:res-FF1-OM-2L}
\mathcal{F}_{1,\text{fin},\text{OM}}^{(2,0,0,\text{re})}
={}& - \frac1{3} L^3 \LEe - L^2 \LEe + L \bigg[ \frac{\LEe^3}{6} - \frac{3}{2}\LEe^2  + \LEe \left( 7 - \frac{4\pi^2}{3} \right) \\
& + 1 + \pi^2 + 4 \zeta_3 \bigg] - \bigg[ \frac{\LEe^4}{24} + \frac{\LEe^3}{2} - \LEe^2 \left( \frac{9}{2} - \frac{7\pi^2}{6} \right) \nonumber \\
& + \LEe \left( 18 - 4\pi^2 + 8 \zeta_3 \right) -3 - \frac{4\pi^2}{3} + \frac{83\pi^4}{180} \nonumber \\
& + \frac{4}{3} \pi^2 \ln^2(2) - \frac{4}{3} \ln^4(2) -32 \Lifoh -18 \zeta_3  \bigg]  \,, \nonumber \\
\mathcal{F}_{1,\text{fin},\text{OM}}^{(2,0,0,\text{im})}
={}& -L^2 \LEe - L \bigg[ \LEe^2 - 6\LEe  - \frac{2\pi^2}{3} \bigg] - \bigg[ \frac{\LEe^3}{3} - \frac{7}{2} \LEe^2 \nonumber \\
&+ \LEe \bigg( 11 + \frac{2\pi^2}{3} \bigg) +1 +\frac{\pi^2}{3} -2 \pi^2 \ln(2) + \zeta_3 \bigg] \,, \nonumber \\
\mathcal{F}_{1,\text{fin},\text{OM}}^{(2,1,0,\text{re})}
={}& -\frac{2}{3} L^2 \LEe - L \bigg[ \frac{4}{3} \LEe^2 - \frac{32}{9} \LEe \bigg] - \bigg[ \frac{4}{9} \LEe^3 - \frac{28}{9} \LEe^2 \nonumber \\
& + \LEe \left( \frac{64}{9} - \frac{10\pi^2}{9} \right) - \frac{32}{9} + \frac{16\pi^2}{27} + \frac{8}{3} \zeta_3  \bigg] \,, \nonumber \\
\mathcal{F}_{1,\text{fin},\text{OM}}^{(2,1,0,\text{im})}
={}& \frac{4}{3} L \LEe + \bigg[ \frac{4}{3}\LEe^2 - \frac{32}{9} \LEe \bigg] \,,\nonumber \\
\mathcal{F}_{1,\text{fin},\text{OM}}^{(2,0,1,\text{re})}
={}& - \frac{2}{3} L^2 \LEe + \frac{32}{9} L \LEe - \bigg[ \LEe \left( \frac{208}{27} - \frac{2\pi^2}{3} \right) + \frac{320}{81} - \frac{8\pi^2}{27} - \frac{8\zeta_3}{3}  \bigg] \,, \nonumber \\
\mathcal{F}_{1,\text{fin},\text{OM}}^{(2,0,1,\text{im})}
={}& \frac{4}{3} L \LEe  - \frac{32}{9} \LEe \nonumber \,.
\end{align}
\endgroup
while
\begin{align}
\mathcal{F}_{1,\text{fin},\text{OM}}^{(2,1,1,\text{re})}=
\mathcal{F}_{1,\text{fin},\text{OM}}^{(2,1,1,\text{im})}=
\mathcal{F}_{1,\text{fin},\text{OM}}^{(2,2,0,\text{re})}=
\mathcal{F}_{1,\text{fin},\text{OM}}^{(2,2,0,\text{im})}=
\mathcal{F}_{1,\text{fin},\text{OM}}^{(2,0,2,\text{re})}=
\mathcal{F}_{1,\text{fin},\text{OM}}^{(2,0,2,\text{im})}
=0\,.
\end{align}
The second form factor reads
\begingroup
\allowdisplaybreaks
\begin{align}
\label{eqn:res-FF2-OM-2L}
\mathcal{F}_{2,\text{fin},\text{OM}}^{(2,0,0,\text{re})}
={}& \frac1{12}L^4 + \frac{5}{6} L^3 + L^2 \bigg[ \frac{3\LEe}{4}-\frac1{2}-\frac{2\pi^2}{3}\bigg]-L\bigg[\LEe(\pi^2-3)+\frac{11}{4}  \\
{}& -\frac{8\pi^2}{3} - 10 \zeta_3 \bigg] - \bigg[ \LEe^2 \left( \frac{\pi^2}{2} -\frac{9}{8} \right) + \LEe \left(\frac{57}{8}-\frac{13\pi^2}{4} -6 \zeta_3  \right) \nonumber \\
{}& -29 + \frac{23\pi^2}{4} + \frac{16\pi^4}{45} - 6 \pi^2 \ln(2) + \frac{4}{3} \pi^2 \ln^2(2) \nonumber \\
{}& - \frac{4}{3}\ln^4(2) - 32\, \text{Li}_4(1/2) + 30 \zeta_3 \bigg] \,,\nonumber \\
\mathcal{F}_{2,\text{fin},\text{OM}}^{(2,0,0,\text{im})}
={}&\frac{2}{3}L^3+L^2\bigg[\frac{\LEe}{2}+\frac{1}{2}\bigg]+L\bigg[\frac{7\LEe}{2}+\pi^2-3\bigg]+\bigg[\frac{3\LEe^2}{2}\nonumber \\
{}& +\LEe\left(\frac{7\pi^2}{6}-8\right) -\frac{13}{4} - \frac{11\pi^2}{3} +2 \pi^2 \ln(2) - 11 \zeta_3 \bigg] \,,  \nonumber \\
\mathcal{F}_{2,\text{fin},\text{OM}}^{(2,1,0,\text{re})}
={}& \frac{4}{9} L^3 +L^2 \bigg[ \frac{2\LEe}{3} + \frac{5}{9} \bigg]+L\bigg[\frac{11\LEe}{3}-\frac{64}{9}+\frac{4\pi^2}{3} \bigg]-\bigg[\frac{7}{36} \LEe^3\nonumber  \\
{}& -\frac{23}{9}\LEe^2 + \LEe \bigg( \frac{649}{54}  - \frac{35\pi^2}{18} \bigg) -\frac{1493}{54} + \frac{122\pi^2}{27} - \frac{4}{3} \zeta_3 \bigg] \,, \nonumber \\
\mathcal{F}_{2,\text{fin},\text{OM}}^{(2,1,0,\text{im})}
={}& \frac{2}{3} L^2 + L \bigg[ 2 \LEe - \frac{50}{9} \bigg] + \bigg[ \frac{3}{2} \LEe^2 - 7 \LEe + \frac{304}{27} -\frac{\pi^2}{3} \bigg] \,, \nonumber \\
\mathcal{F}_{2,\text{fin},\text{OM}}^{(2,0,1,\text{re})}
={}& \frac{4}{9} L^3 + \frac{5}{9} L^2 + L \bigg[ \LEe - \frac{64}{9} + \frac{4\pi^2}{3}  \bigg] - \bigg[ \frac{\LEe^3}{12}  - \frac{\LEe^2}{6} \nonumber  \\
{}&  + \LEe \bigg( \frac{5}{3} - \frac{11\pi^2}{18} \bigg) -\frac{13297}{324} + \frac{145\pi^2}{27} -\frac{4}{3} \zeta_3 \bigg] \,, \nonumber \\
\mathcal{F}_{2,\text{fin},\text{OM}}^{(2,0,1,\text{im})} 
={}&\frac{2}{3} L^2 + L \bigg[ \frac{2\LEe}{3} - \frac{50}{9} \bigg] + \bigg[ \frac{\LEe^2}{6} - \frac{19}{9} \LEe + \frac{304}{27} -\frac{\pi^2}{3} \bigg] \,, \nonumber \\
\mathcal{F}_{2,\text{fin},\text{OM}}^{(2,1,1,\text{re})}
={}& \frac{8}{9} L^2 + L \bigg[ \frac{8}{9} \LEe - \frac{80}{27} \bigg] - \bigg[ \frac{40}{27} \LEe - \frac{200}{81} + \frac{8\pi^2}{9} \bigg] \,,  \nonumber \\
\mathcal{F}_{2,\text{fin},\text{OM}}^{(2,1,1,\text{im})}
={}&  - \frac{16}{9} L - \bigg[ \frac{8}{9} \LEe - \frac{80}{27}  \bigg] \,, \nonumber \\
\mathcal{F}_{2,\text{fin},\text{OM}}^{(2,2,0,\text{re})}
={}& \frac{4}{9} L^2 + L \bigg[ \frac{8}{9} \LEe - \frac{40}{27} \bigg] + \bigg[ \frac{4}{9} \LEe^2 -\frac{40}{27} \LEe + \frac{100}{81} - \frac{4\pi^2}{9} \bigg] \,, \nonumber \\
\mathcal{F}_{2,\text{fin},\text{OM}}^{(2,2,0,\text{im})}
={}& - \frac{8}{9} L - \bigg[ \frac{8}{9}  \LEe -\frac{40}{27}\bigg] \,, \nonumber \\
\mathcal{F}_{2,\text{fin},\text{OM}}^{(2,0,2,\text{re})}
={}& \frac{4}{9} L^2 - \frac{40}{27} L  + \bigg[  \frac{100}{81} - \frac{4\pi^2}{9} \bigg] \,, \nonumber \\
\mathcal{F}_{2,\text{fin},\text{OM}}^{(2,0,2,\text{im})}
={}& - \frac{8}{9} L +  \frac{40}{27} \,. \nonumber
\end{align}
\endgroup

\subsubsection{Equal masses case}
\label{sec:res-EM}
Finally, we present results for equal masses $\me = \mE \ll \sqrt{s}$ and with the scale set to $\mu^2=\mEsq$. At one-loop, we find
\begingroup
\allowdisplaybreaks
\begin{align}
    \label{eqn:res-FF1-EM-1L}
    \mathcal{F}_{1,\text{fin},\text{EM}}^{(1,0,\text{re})} = -L + 2  \,, \qquad
    \mathcal{F}_{1,\text{fin},\text{EM}}^{(1,0,\text{im})} =1 \,, \qquad \mathcal{F}_{1,\text{fin},\text{EM}}^{(1,1,\text{re})} =
     \mathcal{F}_{1,\text{fin},\text{EM}}^{(1,1,\text{im})} = 0\,,
\end{align}
\endgroup
and
\begingroup
\allowdisplaybreaks
\begin{alignat}{4}
    \label{eqn:res-FF2-EM-1L}
    & \mathcal{F}_{2,\text{fin},\text{EM}}^{(1,0,\text{re})} ={} && \frac{1}{2}L^2 +  L + \bigg[ \frac{7\pi^2}{6}  -4   \bigg] \,, \qquad
    &&   \mathcal{F}_{2,\text{fin},\text{EM}}^{(1,0,\text{im})} ={} && L - 1 \,, \\
    & \mathcal{F}_{2,\text{fin},\text{EM}}^{(1,1,\text{re})} ={}&& \frac{2}{3} L - \frac{10}{9} \,, \qquad
    && \mathcal{F}_{2,\text{fin},\text{EM}}^{(1,1,\text{im})} ={} && -\frac{2}{3} \,, \nonumber
\end{alignat}
\endgroup
for the first and second form factor, respectively. At two loops, the form factors read
\begingroup
\allowdisplaybreaks
\begin{align}
    \label{eqn:res-FF1-EM-2L}
    \mathcal{F}_{1,\text{fin},\text{EM}}^{(2,0,\text{re})} ={} &
    -\frac{1}{3}L^3 + L \bigg[ 4 - \frac{2\pi^2}{3}  \bigg]
    + \bigg[ \frac{7\pi^2}{3} - \frac{4\pi^2}{3} \ln(2) -16  \bigg] \,,  \\
    \mathcal{F}_{1,\text{fin},\text{EM}}^{(2,0,\text{im})} ={} & -L^2 + 8L -\bigg[ \frac{34}{3} + \frac{7\pi^2}{9} \bigg]\,, \nonumber \\
    \mathcal{F}_{1,\text{fin},\text{EM}}^{(2,1,\text{re})} ={} & -\frac{2}{3} L^2 + \frac{32}{9} L
    + \bigg[ \frac{2\pi^2}{9} - \frac{16}{9} \bigg] \,, \nonumber \\
    \mathcal{F}_{1,\text{fin},\text{EM}}^{(2,1,\text{im})} ={} & \frac{4}{3} L -\frac{32}{9} \,, \nonumber \\
    \mathcal{F}_{1,\text{fin},\text{EM}}^{(2,2,\text{re})} ={} & \mathcal{F}_{1,\text{fin},\text{EM}}^{(2,2,\text{im})} = 0 \nonumber \,,
\end{align}
\endgroup
and
\begingroup
\allowdisplaybreaks
\begin{align}
\label{eqn:res-FF2-EM-2L}
\mathcal{F}_{2,\text{fin},\text{EM}}^{(2,0,\text{re})} ={} &
\frac{1}{12}L^4 + \frac{1}{2} L^3 - L^2 \bigg[ \frac{1}{2} + \frac{2\pi^2}{3}  \bigg] + L \bigg[ \frac{13\pi^2}{6} + 8 \zeta_3 + \frac{3}{4}  \bigg] \\
{} & + \bigg[ \frac{113\pi^4}{180} +\frac{29\pi^2}{12} - \frac{28\pi^2}{3} \ln(2) -  30\zeta_3 + 11  \bigg]\,, \nonumber  \\
\mathcal{F}_{2,\text{fin},\text{EM}}^{(2,0,\text{im})} ={} & \frac{2}{3} L^3 - \frac{1}{2} L^2 + L \bigg[ \frac{2\pi^2}{3} + 5 \bigg] +\bigg[ \frac{\pi^2}{18} -14 \zeta_3 - \frac{241}{12} \bigg]\,, \nonumber \\
\mathcal{F}_{2,\text{fin},\text{EM}}^{(2,1,\text{re})} ={} & \frac{4}{9} L^3 - \frac{1}{9} L^2  + L \bigg[ \frac{4\pi^2}{3} - \frac{32}{9} \bigg]  -\bigg[ \frac{32\pi^2}{9} -\frac{8}{3}\zeta_3 - \frac{143}{6} \bigg] \,, \nonumber \\
\mathcal{F}_{2,\text{fin},\text{EM}}^{(2,1,\text{im})} ={} & \frac{2}{3}L^2 - \frac{38}{9} L + \bigg[  \frac{208}{27} - \frac{\pi^2}{3}  \bigg] \,, \nonumber \\
\mathcal{F}_{2,\text{fin},\text{EM}}^{(2,2,\text{re})} ={} & \frac{4}{9} L^2 - \frac{40}{27} L - \bigg[ \frac{4\pi^2}{9} - \frac{100}{81} \bigg] \,, \nonumber \\
\mathcal{F}_{2,\text{fin},\text{EM}}^{(2,2,\text{im})} ={} & - \frac{8}{9} L + \frac{40}{27} \nonumber \,.
\end{align}
\endgroup

\section{Conclusion}
\label{sec::conclusion}

In this paper, we have shown how the effective theory 
renormalization-group method of Ref.~\cite{Delto:2025epy} can
be used to extend the classical double-logarithmic analysis of
the high-energy electron-to-muon forward annihilation or,
equivalently, backward scattering~\cite{Gorshkov:1966ht}. The
solution of the resulting two-dimensional partial differential
evolution equation sums up large logarithms and determines the
high-energy behavior of the forward amplitude through the
next-to-leading logarithmic approximation retaining the exact 
dependence on the electron-to-muon mass ratio. The subleading
logarithms  further enhance the cross section relative to the
leading double-logarithmic result. We have also presented a
detailed calculation of the complete two-loop forward amplitude
at leading power in the high-energy limit, considering both the
physical $m_e\ll m_\mu$ and the equal-mass configurations. The
two-loop result provides independent validation of the
renormalization group method while also determining ${\cal O}(\alpha^2)$ 
contributions beyond the next-to-leading logarithmic accuracy.
Our analysis gives an example and a natural starting point
for using the multidimensional Wilsonian renormalization group
flow for the study of the high-energy asymptotic behavior of
amplitudes in various kinematical regimes  at subleading
power.

\vspace{10mm}
\noindent
{\bf Acknowledgments}\\[1mm]
M.D. thanks Philipp B{\"o}er for numerous stimulating discussions.
The work of  A.P. was supported in part by NSERC and the
Perimeter Institute for Theoretical Physics. L.T. and M.D. were
supported in part by the Excellence Cluster ORIGINS funded by
the Deutsche Forschungsgemeinschaft (DFG, German Research
Foundation) under Germany’s Excellence Strategy –
EXC-2094-390783311 and in part by the European Research Council
(ERC) under the European Union’s research and innovation
programme grant agreements 949279 (ERC Starting Grant
HighPHun). The work of M.D. is in part funded by the
European Union (ERC, grant agreement No. 101044599).

\appendix

\section{Asymptotic behavior of $k(z)$}
\label{app:asyexp-k}

In this appendix we derive the asymptotic behavior of the function
\begin{align}
k(z) = z\int_{0}^1{\rm d}\eta'\int_{\eta'}^1{\rm d}\xi'
(\xi'-\eta')\ALL(1-\eta',1-\xi')\ALL(\xi',\eta')
\,,
\end{align}
see \cref{eq::kz}, for $z \to \infty$. The form factor $\ALL$ was defined in \cref{eq::Ares}. First, we use the asymptotic behavior of Bessel functions for large arguments
\begin{align}
    I_n(z) \underset{z \to \infty}{\sim} \frac{e^z}{\sqrt{2\pi z}} \,,
\end{align}
and write
\begin{align}
    k(z) \underset{z \to \infty}{\sim} \frac{\sqrt{z}}{4\pi} \int_{0}^1 {\rm d} \eta' \, \int_{\eta'}^1 {\rm d} \xi' \, P(\eta',\xi') \, e^{2\sqrt{z}g(\eta',\xi')} \,,
\end{align}
where
\begin{align}
    g(\eta',\xi') ={} & \sqrt{(1-\eta')(1-\xi')} + \sqrt{\eta'\xi'} \,, \\
    P(\eta',\xi') ={} & \frac{(\xi'-\eta')^3}{\xi'(1-\eta')[(1-\eta')(1-\xi')\eta'\xi']^{1/4}} \,.
\end{align}
We find it simpler to perform a saddle-point analysis  for large $z$ asymptotics after decoupling integrations. By writing $\{\eta'=x,\xi' = x +(1-x)y \}$, we arrive at
\begin{align}
    k(z)\underset{z \to \infty}{\sim}
    \frac{\sqrt{z}}{4\pi}
    \int_{0}^1 {\rm d}x \int_0^1 {\rm d}y \, P(x,y) \, e^{2\sqrt{z}g(x,y)} \,,
    \label{eqn:koz-asy-xy}
\end{align}
where
\begin{align}
    g(x,y) ={} & (1-x) \sqrt{(1-y)} + \sqrt{x(x + (1-x)y)} \,, \\
    P(x,y) ={} & \frac{(1-x)^3 y^3}{[x + (1-x)y][(1-x)^2x(1-y)(x + (1-x)y)]^{1/4}} \,.
\end{align}
The integral in \cref{eqn:koz-asy-xy} is dominated by the region $y\to 0$, and we approximate
\begin{align}
    g(x,y) ={} & 1 - \frac{1-x}{8x} y^2 + \ORD{y^3} \,, \\
    P(x,y) ={} & \frac{(1-x)^{5/2}}{x^{3/2}}y^3 + \ORD{y^4} \,,
\end{align}
such that after integration over $y$, 
\begin{align}
    \label{eqn:koz-asy-x}
    k(z)={} & 
    \frac{e^{2\sqrt{z}}}{2\pi \sqrt{z}}
    \int_{0}^1 {\rm d}x \, 4 \sqrt{x(1-x)} \left[1 - \exp\!\left( -\frac{(1-x)\sqrt{z}}{4x} \right) \left( 1 + \frac{(1-x)\sqrt{z}}{4x} \right)  \right] \\
    {}& \times \left[1 + \ORD{z^{-1/4}}\right]  \,. \nonumber
\end{align}
The second term in the square brackets is power suppressed for $z\to\infty$, as it vanishes on $x\in[0,1)$. We drop it and arrive at
\begin{align}
    k(z) = 
    \frac{e^{2\sqrt{z}}}{2\pi \sqrt{z}}
    \int_{0}^1 {\rm d}x \, 4 \sqrt{x(1-x)} \times \left[1 + \ORD{z^{-1/4}}\right]  = \frac{e^{2\sqrt{z}}}{4\sqrt{z}} \times \left[1 + \ORD{z^{-1/4}}\right] \,.
    \label{eqn:koz-asy-final}
\end{align}

\section{Renormalization constants}
\label{app:renconst}
Renormalized form factors can be obtained from their bare counterparts through
\begin{align}
    \label{eqn:FFren-from-bare-0L}
    \mathcal{F}_{i,\mathrm{ren}}^{(0)} ={} & \mathcal{F}_{i,\mathrm{b}}^{(0)} = \delta_{i2} \,,  \\
    \label{eqn:FFren-from-bare-1L}
    \mathcal{F}_{i,\mathrm{ren}}^{(1)} ={} & \mathcal{F}_{i,\mathrm{b}}^{(1),0} + \mathcal{F}_{i,\mathrm{b}}^{(0)} \left[ \delta C_{\overline{\alpha}\alpha}^{(1)} + \delta Z_{\alpha}^{\overline{\text{MS}},(1)} + \delta Z_{2,e}^{\text{OS},(1)} + \delta Z_{2,\mu}^{\text{OS},(1)} \right] \,,  \\
    \label{eqn:FFren-from-bare-2L}
    \mathcal{F}_{i,\mathrm{ren}}^{(2)} ={} & \mathcal{F}_{i,\mathrm{b}}^{(2)}
    + \mathcal{F}_{i,\mathrm{b}}^{(1),0} \left[ 2 \delta C_{\overline{\alpha}\alpha}^{(1)} + 2 \delta Z_{\alpha}^{\overline{\text{MS}},(1)} + \delta Z_{2,e}^{\text{OS},(1)} + \delta Z_{2,\mu}^{\text{OS},(1)} \right]  \nonumber \\
    & + \mathcal{F}_{i,\mathrm{b}}^{(1),\me}  \delta Z_{\me}^{\text{OS},(1)}  + \mathcal{F}_{i,\mathrm{b}}^{(1),\mE} \delta Z_{\mE}^{\text{OS},(1)} \nonumber  \\
    & + \mathcal{F}_{i,\mathrm{b}}^{(0)} \bigg[  \delta C_{\overline{\alpha}\alpha}^{(2)} + \delta Z_{\alpha}^{\overline{\text{MS}},(2)} + \delta Z_{2,e}^{\text{OS},(2)} + \delta Z_{2,\mu}^{\text{OS},(2)}   +  \delta Z_{2,e}^{\text{OS},(1)} \delta Z_{2,\mu}^{\text{OS},(1)} \nonumber  \\
    & + 2 \delta C_{\overline{\alpha}\alpha}^{(1)} \delta Z_{\alpha}^{\overline{\text{MS}},(1)} + \left( 2 \delta C_{\overline{\alpha}\alpha}^{(1)} + \delta Z_{\alpha}^{\overline{\text{MS}},(1)} \right) \left( \delta Z_{2,e}^{\text{OS},(1)} + \delta Z_{2,\mu}^{\text{OS},(1)} \right)  \bigg] \,,
\end{align}
where coefficients of the loop expansion of the renormalization constants were defined in \cref{eqn:renconst-exp-general}. The coefficients of the QED coupling renormalization constant in $\overline{\text{MS}}$ read
\begin{align}
    \delta Z_{\alpha}^{\overline{\text{MS}},(1)} = \frac{2(\ne + \nE)}{3\epsilon}  \,, \qquad
    \delta Z_{\alpha}^{\overline{\text{MS}},(2)} = \frac{4(\ne + \nE)^2}{9\epsilon^2}  + \frac{(\ne + \nE)}{2\epsilon}  \,,
    \label{eq:Zalpha}
\end{align}
and those of the on-shell scheme conversion factor, taken from \cite{Sturm:2013uka}, read
\begin{align}
    \delta C_{\overline{\alpha}\alpha}^{(1)} =  \frac{2}{3} \sum_{F=e,\mu} \nF \, l_F  \,, \qquad
    \delta C_{\overline{\alpha}\alpha}^{(2)} =   \sum_{F=e,\mu} \nF \left[   \left( \frac{15}{4} + l_F \right) + \frac{4}{9} l_F \sum_{F'=e,\mu} \nFp \, l_{F'}  \right] \,,
    \label{eq:Caa}
\end{align}
where $l_F=\ln(\mu^2/m_F^2)$ with the mass renormalized in the on-shell scheme. We obtain the wave-function renormalization by abelianizing the results in Refs.~\cite{Bekavac:2007tk,Fael:2020bgs}; for $F={e,\mu}$, we write
\begin{align}
    \delta Z_{2,F}^{\text{OS},(1)} ={} & \bigg[ -\frac{3}{2\epsilon} - \left( 2 + \frac{3}{2} l_{F} \right)
    - \epsilon \left( \frac{\pi^2}{8} + 4 + 2 l_F +\frac{3}{4} l_F^2 \right) \nonumber \\
    {}& + \epsilon^2 \bigg( \frac{\zeta_3}{2} - \frac{\pi^2}{6} - 8 - 4 l_F 
    -\frac{\pi^2}{8} l_F - l_F^2 - \frac1{4} l_F^3 \bigg) + \ORD{\epsilon^3} \bigg] \,, \\
    \delta Z_{2,F}^{\text{OS},(2)} ={} & \bigg[ \frac{9}{8\epsilon^2}
    + \frac1{\epsilon} \left( \frac{51}{16} + \frac{9}{4} l_F \right) 
    + \bigg( \frac{9}{4} l_F^2 + \frac{51l_F}{8} + 4 \pi^2 \ln(2) 
    - \frac{49\pi^2}{16} + \frac{433}{32}  \nonumber \\
    {} & - 6\zeta_3 \bigg) \bigg]  + \nF \bigg[ \frac1{\epsilon} \bigg( \frac{1}{4} + l_F \bigg) 
    + \bigg( \frac{947}{72} - \frac{5 \pi^2}{4} + \frac{11}{6} l_F + \frac{3}{2} l_F^2 \bigg) \bigg] \nonumber \\
    {} & + \nFb \bigg[ \frac1{\epsilon} \bigg( \frac{1}{4} + l_F -2 \ln(r)  \bigg)
    + \bigg( \frac{3}{2} l_F^2 + \frac{11 l_F}{6} - 4 l_F \ln(r)  \nonumber \\
    {} & + \big(\frac{8}{3}+4r^2 \big) \ln(r) + \big(4+6r^4\big) \ln^2(r) 
    \nonumber \\
    {}& + \big(5 r^3+3 r-2-6 r^4 \big) \big(\ln(1-r)\ln(r) + \Li{2}{r} \big) \nonumber \\
    {} & - \big(6 r^4+5 r^3+3 r+2 \big) \big(\ln(1+r)\ln(r) + \Li{2}{-r} \big) \nonumber \\
    {} & + \big( \frac{5}{12} - \frac{3}{2}r -\frac{5}{2}r^3 + r^4 \big) \pi^2 + \frac{443}{72} + 7 r^2 \bigg) \bigg] + \ORD{\epsilon} \,,
\end{align}
where $r=m_{\overline{F}}/m_F$ and with all logarithms and polylogarithms evaluated on their principal branches. Note that starting from the two-loop coefficient $\delta Z_{2,F}^{\text{OS},(2)}$, the wave function renormalization constant receives contributions from the other flavor $\overline{F}={\mu,e}$ running in the loop. Finally, mass-renormalization constants are only required to one-loop order, for which they are identical to the wave-function renormalization constant $\delta Z_{m_F}^{\text{OS},(1)} =  \delta Z_{2,F}^{\text{OS},(1)}$.

\bibliographystyle{JHEP}
\bibliography{biblio}

\end{document}